\documentclass[10pt]{article}

\usepackage[letterpaper,margin=1in]{geometry}
\usepackage[T1]{fontenc}
\usepackage{times}
\usepackage{amsmath}
\usepackage{amssymb}
\usepackage{booktabs}
\usepackage{graphicx}
\usepackage{microtype}
\usepackage[hidelinks]{hyperref}
\usepackage{natbib}
\bibpunct{(}{)}{;}{a}{}{,}
\usepackage{caption}
\graphicspath{{figures/}}

\newcommand{\prov}[1]{#1$^{\dagger}$}
\newcommand{\notejustify}{\par\leftskip=0pt\rightskip=0pt\parfillskip=0pt plus 1fil\relax}
\newcommand{\tablenotes}[1]{\par\smallskip{\footnotesize\textit{Note.} #1\par}}

\title{\vspace{-2.5em}Lomb-Scargle Recovery of Periodic Binary Signals across Black Hole Mass and Cadence}

\author{
  Karan Akbari\textsuperscript{1}\thanks{Corresponding author: \texttt{karanakbari14@gmail.com}; \href{https://orcid.org/0009-0005-0550-4018}{ORCID 0009-0005-0550-4018}}
  \ and Sayantan Bhattacharya\textsuperscript{2}\thanks{\texttt{sayantan.bhattacharya@tifr.res.in}; \href{https://orcid.org/0000-0001-8572-8241}{ORCID 0000-0001-8572-8241}}\\[6pt]
  \parbox{0.9\textwidth}{\centering\small \textsuperscript{1}St.~Xavier's College, Mumbai, India\\
  \textsuperscript{2}Department of Astronomy and Astrophysics, Tata Institute of Fundamental Research, 1 Homi Bhabha Road, Colaba, Mumbai, India}
}
\date{}

\begin{document}
\maketitle

\begin{abstract}
Searches for periodic supermassive black hole binaries in quasar light curves rely on recovery fractions, the share of injected signals detected, which forecasts use as detection efficiencies. \citet*{Lin2026} measured them for the Lomb-Scargle periodogram on damped-random-walk light curves matched to 12,400 bright, high-mass Palomar Transient Factory (PTF) quasars with Sloan Digital Sky Survey Data Release 7 (DR7) masses. To test them over four decades of mass, we inject their sinusoidal and sawtooth signals into light curves of mergers in the TNG100-1 cosmological simulation, with magnitudes drawn from DR7 quasars of similar mass (extrapolated at low mass). We apply their threshold in each frequency bin, set by $10^5$ noise-only light curves per binary, on observed PTF cadences, one Stripe~82 cadence and an LSST-like cadence (Legacy Survey of Space and Time). At fixed absolute magnitude $M_i=-23$, sinusoid recovery falls toward high mass, and with DR7 magnitudes it climbs above about $10^8\,M_\odot$, so the magnitude assignment sets the sign of the trend. Above $10^8\,M_\odot$, where their sample lies, PTF sinusoid recovery is 39.8 against their 44.7 per cent. At LSST-like cadence sinusoid recovery is 76.9 per cent, about three times their 22.6 per cent, for reasons mostly unidentified. Denser sampling of the same PTF baselines did not lower recovery in any test. Even at LSST-like cadence we miss most sawtooth signals above $10^8\,M_\odot$. We tabulate recovery per 0.5-dex mass bin and cadence, with low-mass values likely too high for a deeper survey.
\end{abstract}

\noindent{\small \textit{Keywords:} Supermassive black holes (1663), Double quasars (406), Active galactic nuclei (16), Quasars (1319), Time domain astronomy (2109), Gravitational wave sources (677)\par}

\section{Introduction}
\label{sec:intro}

Supermassive black hole binaries form when two galaxies merge and their central black holes sink to a shared centre \citep{Begelman1980}. The subparsec pairs that survive that inspiral are the loudest individual sources in the nanohertz gravitational-wave band, and pulsar timing arrays have now measured a stochastic background consistent with a cosmological population of them \citep{Agazie2023evidence,Agazie2023smbhb}. A confirmed electromagnetic counterpart would tie one such binary to that background and fix its mass, redshift, and orbital period \citep{Bogdanovic2022}, so the counterpart hunt feeds both the timing-array and the accretion-physics sides of the problem.

Most counterpart searches look for periodic modulation in the optical light curves of quasars \citep{Graham2015,Charisi2016,Liu2019,Chen2020,Charisi2022}. A binary drives the accretion rate onto the individual holes at the orbital period, or at a period a few times longer through the overdense lump the circumbinary disk builds, and hydrodynamic simulations of that flow predict quasi-periodic optical variability \citep{MacFadyen2008,DOrazio2013,Farris2014,Tang2018}. \citet{Graham2015} reported 111 candidates in the Catalina Real-Time Transient Survey (CRTS), and \citet{Charisi2016} found 50 significantly periodic quasars in the Palomar Transient Factory (PTF) database, 33 of which held up once extra data from iPTF, the successor of PTF, and from CRTS were folded in.

The trouble is the background: a quasar varies as a damped random walk (DRW) to good approximation \citep{Kelly2009,MacLeod2010,Koz2010}, and that red-noise power spectrum falls as $f^{-2}$ at high frequency, so a few cycles of apparent periodicity turn up in purely stochastic data with high probability \citep{Vaughan2016}. Any periodicity claim rests on a null model that carries this red noise, whether the test is a periodogram threshold or Bayesian model selection against a DRW \citep{Witt2022}. \citet{Robnik2024} reran the \citet{Charisi2016} sample with a test whose frequentist properties hold whatever the noise distribution is, so no noise simulation is needed, and none of the 33 survived. A systematic search of Zwicky Transient Facility quasars found the candidate modulation quasi-periodic at best, with only a few surviving a longer baseline \citep{ChenYJ2024}.

\citet*[hereafter LCH26]{Lin2026} took the significance calibration as given and asked a different question. When a real periodic signal is present, how often does the Lomb-Scargle periodogram (LSP) recover it at all? LCH26 injected sinusoidal and sawtooth signals (a linear ramp that resets once per period) into synthetic DRW light curves matched to 12,400 spectroscopically confirmed PTF quasars with at least 50 observed nights and a virial black hole mass in the Sloan Digital Sky Survey (SDSS) Data Release 7 (DR7), and measured the recovery fraction, the share of injected signals the LSP detects, on three cadences: PTF-like (one epoch per observed night of the PTF light curves), idealised (one epoch per day over the same baselines) and LSST-like (modelled on the Legacy Survey of Space and Time, LSST, of the Vera C.\ Rubin Observatory; \citealt{Ivezic2019}). The recovery fraction is low: a maximum of 44.7 per cent of injected sinusoids and only up to 9.3 per cent of sawtooth signals are recovered in their PTF-like light curves, and the idealised and LSST-like cadences do worse. LCH26 conclude that earlier searches ``must have missed a significant fraction of periodic signals'', with the miss close to total for sawtooth-shaped modulation. \citet{Bertassi2026} analysed the same light curves with a Gaussian-process model with a periodic kernel and found higher recovery on the denser cadences.

Forecasts of how many binaries optical surveys will find use recovery fractions as detection efficiencies \citep[e.g.,][]{Kelley2019,Akbari2026}, and the LCH26 fractions are tied to one mass and one luminosity distribution. LCH26 read that distribution as broad (their Figure~1 caption calls the 12,400 ``representative of the entire quasar population''), but a spectroscopic PTF-plus-DR7 selection reaches the bright, high-mass end of the quasar population and very little of the binary mass distribution below it, so how far the recovery fractions travel across that distribution is not known.

The coming decade of dense optical monitoring, above all LSST, makes the mass reach of these searches a practical question. Mass-binned detectability of simulated massive binaries has precedent \citep{Kelley2019,Chiesa2026}. We run LSP injection-recovery on the TNG100-1 black hole merger catalogue of IllustrisTNG \citep{Springel2018,Pillepich2018,Nelson2019}, which supplies 18,374 merging black hole pairs with known primary mass and redshift over four decades of mass, free in mass of the magnitude cut that shapes DR7. We put the sinusoidal and sawtooth signals side by side, apply the LCH26 detection threshold, set in each frequency bin from noise-only DRW light curves, draw an $i$-band magnitude for each binary from DR7 quasars of similar mass, and use archival cadences (the observed PTF cadences of the \citet{Charisi2016} quasars and one observed Stripe~82 cadence) and an LSST-like cadence built to the LCH26 specification (Section~\ref{sec:cadences}). We also test each input in which their text leaves a choice open or our setup differs from theirs. The first question is how recovery depends on black hole mass and on how the magnitude is assigned. The second is whether the published PTF fractions are reproduced, which a search of the PTF archive takes for granted when it adopts them as its detection efficiency. Our LSST-like value is then compared with theirs, because a forecast of the binaries LSST will find multiplies by the LSST-like recovery fraction, and theirs is the only published periodogram recovery fraction for a full quasar sample at that cadence. At fixed baseline we test whether denser sampling lowers recovery. The recovery fractions in 0.5-dex mass bins, for both signal shapes on the three types of cadence, let a forecast use a detection efficiency for each mass instead of one published value for all masses. Section~\ref{sec:population} describes the simulated binaries, their magnitudes and the cadences, Section~\ref{sec:methods} the light curves and the detection rule, Sections~\ref{sec:results-ptf-s82} and~\ref{sec:results-lsst} the results, and Sections~\ref{sec:discussion} and~\ref{sec:conclusions} the discussion and conclusions. The appendices hold the checks and the list of PTF cadences.

\section{Sample and Data}
\label{sec:population}

\subsection{Simulated binaries}
\label{sec:binaries}

The TNG100-1 run \citep{Nelson2019} records every black hole merger in the box, 18,374 events in its public merger catalogue, and we take each merger to stand for a binary (Section~\ref{sec:limitations}). No event falls below a total mass of $1.2\times10^6\,M_\odot$, the black hole seed mass of the simulation, below which a merger would be a numerical artefact (the smallest total is $2.36\times10^6\,M_\odot$). Primary masses come from the black hole details record, which stores each black hole's mass at every time step, and not from the merger catalogue, which notes that the mass it lists for the surviving black hole is that of the gas cell holding it. We take as the primary mass $M_{\rm BH,1}$ the larger of the surviving black hole's mass in its last details entry before the merger time and the infalling black hole's mass as listed in the merger catalogue, converted to solar masses with the simulation's $h=0.6774$, and write $\log M_1$ for $\log_{10}(M_{\rm BH,1}/M_\odot)$. The primary masses span $\log M_1\in[6.07,10.09]$ with median 7.87, and 46.4 per cent sit at $\log M_1\ge8$. Redshifts, $z=1/a-1$ at the scale factor $a$ of the merger, run out to $z=7.24$ with median 0.95. Of the simulation's quantities, only the primary mass and the redshift enter the light-curve model. We call each simulated binary, with its primary mass, redshift and the magnitude and light curves we give it, a system. The 65 systems with $\log M_1\ge10$ lie outside the mass bins of Section~\ref{sec:sampling} and are never drawn.

\subsection{Quasar magnitudes}
\label{sec:magnitudes}

The DRW amplitude depends on the absolute $i$-band magnitude as well as the black hole mass (Section~\ref{sec:drw}), so every system needs one. The magnitude is $M_i(z=2)$ throughout, the SDSS convention of an absolute $i$-band magnitude $K$-corrected to $z=2$, which is what the \citet{MacLeod2010} relations and LCH26 both take; we write it $M_i$ from here on. The obvious choice, the luminosity TNG100-1 assigns from the accretion rate at the merger, is wrong for this purpose. Most TNG100-1 binaries are not luminous quasars when they merge, and their $M_i$ at the merger has a median near $-15.7$, far below the quasar range LCH26 draw from. A single fixed $M_i$ for all systems is wrong too, because $M_i$ and black hole mass are correlated in any real quasar sample, and a fixed $M_i$ removes the part of the mass dependence that this correlation carries into the noise.

\begin{figure}[tbp]
\centering
\includegraphics[width=0.6\textwidth]{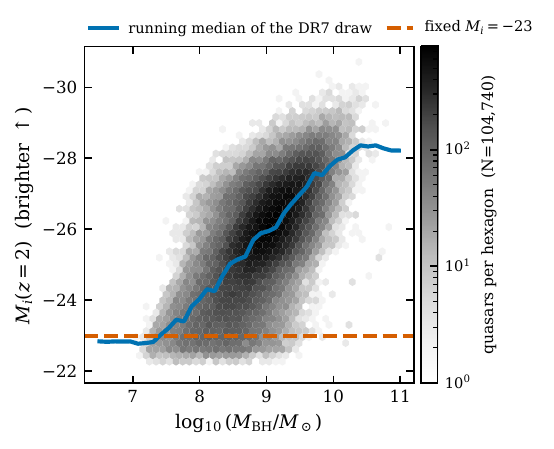}
\caption{Absolute magnitude $M_i(z=2)$ against black hole mass for the 104,740 \citet{Shen2011} DR7 quasars from which magnitudes are drawn (grey hexagons, shaded by count). The solid curve is the running median of the drawn magnitudes and the dashed line the single $M_i=-23$ given to every system in two further sets of injections (Section~\ref{sec:sampling}). The magnitude axis is inverted so that brighter quasars sit higher.}
\label{fig:mi_mechanism}
\end{figure}

We assign $M_i$ by a conditional draw from the \citet{Shen2011} catalogue of SDSS DR7 quasars \citep{abazajian2009}, using its fiducial virial black hole mass and its $M_i(z=2)$. With the DR7 objects sorted by mass, a binary of mass $M_1$ takes the $M_i$ of one object picked uniformly at random from the 200 nearest it in that ordering (100 on each side, shifted inward at the two ends of the catalogue), so the joint distribution of mass and magnitude that LCH26 inherit from DR7 is reproduced on the TNG100-1 masses (Figure~\ref{fig:mi_mechanism}). The draw is independent for each binary and uses mass only, so one DR7 object can supply several binaries and $M_i$ does not depend on redshift. We made the draw once, with a fixed seed (Appendix~\ref{app:draws}), and use the same magnitudes throughout. The catalogue carries 105,783 quasars, and 104,740 of them have a virial mass estimate below $10^{11}\,M_\odot$ (1037 have none and 6 lie above), all with an $M_i$, which is the set we draw from; its median is $\log_{10}(M_{\rm BH}/M_\odot)=9.0$.

The DR7 catalogue thins out and then stops at the low-mass end, with 3 objects in $[6.5,7.0)$, 473 in $[7.0,7.5)$ and none below $\log_{10}(M_{\rm BH}/M_\odot)=6.502$. Below that mass the draw falls back to a linear extrapolation of the global relation, a straight-line fit to all 104,740 objects, $M_i=-9.627-1.808\,\log_{10}(M_{\rm BH}/M_\odot)$, with the same fitted residual scatter (Gaussian, standard deviation 1.020\,mag), so the magnitudes in the lowest mass bin are extrapolations, with no DR7 objects underneath to draw from. The 2974 systems that fall there are 16.2 per cent of the 18,374. Every mass from 6.502 to about 7.30 pulls the same 200 neighbours, the lowest 200 DR7 objects (median $M_i=-22.82$), so the second mass bin and the lower part of the third share a single magnitude distribution at different masses. The running median of the drawn magnitudes brightens from $M_i=-20.6$ at the lowest TNG100-1 mass to $-28.0$ at the highest.

We call $\log M_1\in[8,10)$ the high-mass range, since it is where the LCH26 sample sits (their 12,400 quasars peak near $\log_{10}(M_{\rm BH}/M_\odot)\approx9$; their Figure~1), and it carries every comparison against them in this paper. The DR7 masses are virial estimates and the TNG100-1 masses are true, so the draw matches a virial relation onto true masses, a mismatch we return to in Section~\ref{sec:limitations}.

\subsection{Cadences}
\label{sec:cadences}

\begin{figure}[tbp]
\centering
\includegraphics[width=0.6\textwidth]{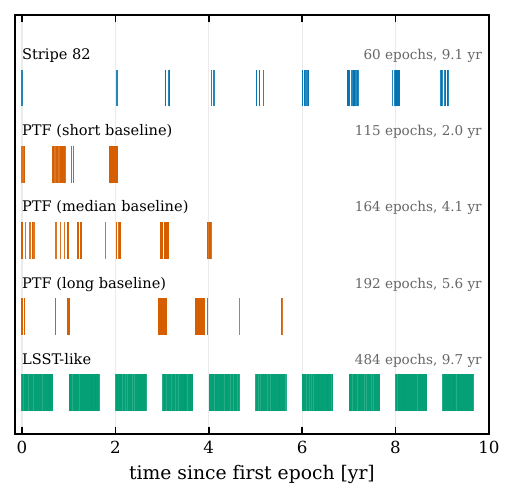}
\caption{Observation times of the cadences against time since the first epoch: the Stripe~82 cadence, three of the 49 PTF cadences (a short-, the median- and a long-baseline example) and the LSST-like cadence. Labels give the number of epochs and the baseline. All three PTF examples hold more epochs than the median PTF cadence (114).}
\label{fig:cadences}
\end{figure}

Three types of cadence are used, a cadence being the list of observation times (epochs) of one light curve (Figure~\ref{fig:cadences}), and we take a night as the integer part of the modified Julian date (MJD) of an epoch. The PTF cadences are the observed timestamps of the \citet{Charisi2016} candidate sample, pulled from the PTF archive in the R band: all good-quality epochs of the 50 quasars in Table~1 of \citet{Charisi2016}, with repeat exposures within a night kept (Appendix~\ref{app:ptfquery} gives the query). We exclude TEX 1428+370, whose 828 epochs on 337 nights are about three times the epochs of the densest other light curve (285). The 49 PTF cadences we use, listed in Table~\ref{tab:ptfcadences}, hold 6207 epochs on 2816 nights, a mean of 127 epochs (median 114) over a mean baseline of 3.9\,yr (median 4.1\,yr), and each system draws one of them (Section~\ref{sec:sampling}). The 49 cadences are longer than the LCH26 PTF-like light curves, which average 3.2\,yr, and sparser in nights. Our records keep intra-night repeat exposures, a mean of 127 epochs spread over 57 distinct nights, while LCH26 sample a one-day grid at one time per observed night of the \citet{Charisi2016} light curves, so a night can contribute at most one epoch on their side.

Stripe~82 is an SDSS field imaged repeatedly over about ten years \citep{MacLeod2010}. The Stripe~82 cadence is the 60 $r$-band observation times of one quasar (Stripe~82 database ID 133735, RA $3.7355^\circ$, Dec $-0.0189^\circ$, $z=2.64$) in the public recalibrated Stripe~82 light curves of the 9258 spectroscopically confirmed DR7 quasars modelled by \citet{MacLeod2010}, chosen from 200 of those light curves drawn at random as the one closest to 60 epochs over 9\,yr. The 60 nights run from MJD 51075.31 to 54412.24, a 9.14-yr baseline with the strong annual gaps of a northern ground-based survey. The Stripe~82 cadence matches none of the LCH26 cadences and stands as an independent archival test, and every system uses this one cadence on Stripe~82.

The LSST-like cadence is our realisation of the LCH26 specification, a 10-yr baseline at 5-day sampling jittered by 1\,d with a four-month annual gap. We place a nominal epoch every 5.0\,d from MJD 60000 over 10\,yr, keep the nominal epochs in the first 243.5\,d of each 365.25-d year (a gap of $4\times30.44$\,d at the end of each year), then shift each kept epoch by a Gaussian jitter of standard deviation 1\,d, which does not accumulate, and sort the times. The last year ends in a gap, so the 484 epochs span 9.66\,yr, and all systems share this one cadence.

\section{Methods}
\label{sec:methods}

\subsection{Damped random walk light curves}
\label{sec:drw}

We build each light curve and its detection test as LCH26 do wherever their text is explicit, except for the inputs listed in Table~\ref{tab:inputs}. Each system's variability is a damped random walk with rest-frame (subscript RF) damping time $\tau$ in days and stationary amplitude $\sigma$ in magnitudes, set by the \citet{MacLeod2010} scaling relations (their Eq.~(7) with the recommended coefficients of their Table~1, reproduced as Eqs.~(2) and~(3) of LCH26),
\begin{align}
\log_{10}(\tau_{\rm RF}) &= 2.4
   + 0.17\,\log_{10}\!\left(\frac{\lambda_{\rm RF}}{4000\,\textrm{\AA}}\right)
   \nonumber\\
   &\; + 0.03\,(M_i + 23)
   + 0.21\,\log_{10}\!\left(\frac{M_{\rm BH}}{10^9\,M_\odot}\right),
   \label{eq:tau}\\
\log_{10}(\sqrt{2}\,\sigma) &= -0.51
   - 0.479\,\log_{10}\!\left(\frac{\lambda_{\rm RF}}{4000\,\textrm{\AA}}\right)
   \nonumber\\
   &\; + 0.131\,(M_i + 23)
   + 0.18\,\log_{10}\!\left(\frac{M_{\rm BH}}{10^9\,M_\odot}\right),
   \label{eq:sigma}
\end{align}
with $M_{\rm BH}=M_{\rm BH,1}$, the primary mass, and no intrinsic scatter about the relations, evaluated at the rest-frame R-band wavelength $\lambda_{\rm RF}=6516.05\,{\rm \AA}/(1+z)$ per system on every cadence (PTF, Stripe~82 and LSST-like), which gives 3341.6\,\AA\ at the median redshift $z=0.95$. LCH26 (their \S2.2.1) use ``the effective wavelength of the R band \ldots\ in the rest frame of the source''. We read that as the redshift-dependent wavelength above. The same wording admits a fixed 6516.05\,\AA\ for every quasar, which would lower $\sigma$ by a factor of about 1.4 and lengthen $\tau$ by about 12 per cent (for a system at $\log M_1=8.5$ with $M_i=-25$ at the median redshift, $\sigma$ falls from 0.106 to 0.077\,mag). We adopt the redshift-dependent reading; we did not run the other, and Table~\ref{tab:inputs} gives its expected effect on recovery.

The damping time enters as the MacLeod rest-frame value, and we use it directly in the recursion below, on observed-frame epoch spacings, without the factor $(1+z)$ that converts a rest-frame time to the observed frame. LCH26 do not state their convention, and their Figure~3 matches the converted value (Appendix~\ref{app:figures}); Appendix~\ref{app:tauz} measures the effect of converting. We take $\tau$ and $\sigma$ from the population scaling and do not fit them per system. A reliable DRW fit needs a baseline of order ten damping times \citep{Kozlowski2017}, longer than the PTF records supply, and the Stripe~82 SDSS baseline alone underconstrains $\tau$ until Pan-STARRS extends it \citep{Suberlak2021}.

We sample the walk at the cadence epochs with the \citet{Kelly2009} Ornstein-Uhlenbeck recursion, starting from a stationary draw $x(t_0)\sim\mathcal{N}(0,\sigma^2)$,
\begin{equation}
\begin{split}
x(t_{i+1}) &= x(t_i)\,e^{-\Delta t/\tau} + \varepsilon, \\
\varepsilon &\sim \mathcal{N}\!\left(0,\,\sigma^2\,(1 - e^{-2\Delta
t/\tau})\right),
\end{split}
\label{eq:ou}
\end{equation}
in the MacLeod convention, where $\sigma$ is the stationary amplitude, $\mathrm{Var}(x)=\sigma^2$ and the structure function at long time lags is $\mathrm{SF}_\infty=\sqrt{2}\,\sigma$. The recursion is linear in epoch count, which keeps the dense LSST-like cadence cheap, and it reproduces a Cholesky draw on the DRW autocovariance (a direct draw from the joint normal distribution of all epochs).

\subsection{Injected signals}
\label{sec:signals}

Onto each light curve we add one of two signals from LCH26, in magnitudes,
\begin{align}
s_{\rm sin}(t)  &= A\,\sin\!\left(2\pi t / P + \phi_0\right), \label{eq:sin}\\
s_{\rm saw}(t)  &= A\left[1 - 2\left\{\frac{t}{P} + \frac{\phi_0}{2\pi}\right\}\right],
\label{eq:saw}
\end{align}
where $\{\cdot\}$ is the fractional part, with the sawtooth giving the sharp-rise, linear-fall shape in magnitudes: at phase zero it jumps $A$\,mag fainter and then brightens linearly over one period to $A$\,mag brighter, so in flux it is a sudden drop followed by a linear rise. Amplitude is drawn $A\sim U[0,0.5)$\,mag and phase $\phi_0\sim U[0,2\pi)$, one injection per system per signal type, with independent amplitudes, periods and phases and independent DRW and noise realisations. Photometric noise is Gaussian per epoch, 0.05\,mag on the PTF cadences, 0.02\,mag on Stripe~82, and 0.01\,mag on the LSST-like cadence. The 0.05 and the 0.01 are both LCH26 values, their stated PTF sample mean and their LSST-like specification, and only the 0.02 on Stripe~82 is ours.

The injected period, in the observed frame, is where our setup and the LCH26 setup part. With $T$ the baseline of a light curve (last minus first epoch), we draw $P\sim U[30,779]$\,d on the PTF and LSST-like cadences, with the upper edge 779\,d equal to $T/1.5$ for the LCH26 mean PTF baseline of 3.2\,yr, though our own PTF cadences run longer, and $P\sim U[200,1100]$\,d on the sparser, longer Stripe~82 cadence. LCH26 instead redraw $P$ for every light curve from $U[0,T/1.5]$ with no lower cutoff; their 30\,d is only the shortest period on the search grid, and their footnote~3 reuses the PTF-like periods for the LSST-like set. The two period distributions are not identical, and Section~\ref{sec:periods} takes up what that difference can and cannot change.

\subsection{Periodogram and detection threshold}
\label{sec:detection}

We compute the LSP with the \texttt{LombScargle} class of \texttt{astropy} 7.2.0 \citep{astropy2022}, weighting by the per-epoch errors, with the data centred, the mean fitted, one Fourier term and the standard normalisation of \texttt{astropy}. A periodogram weighted by the errors and with a fitted mean is the generalised periodogram, which the \texttt{astroML} routine that LCH26 use \citep{VanderPlas2012} also computes. The frequency grid is the LCH26 one, 1000 frequencies spaced evenly in $\log f$ between $1.5/T$ and $1/30$ cycles per day, both ends included, and we split it into 25 bins $B_k$ of equal width in $\log f$, with 40 frequencies each.

\begin{figure}[tbp]
\centering
\includegraphics[width=\textwidth]{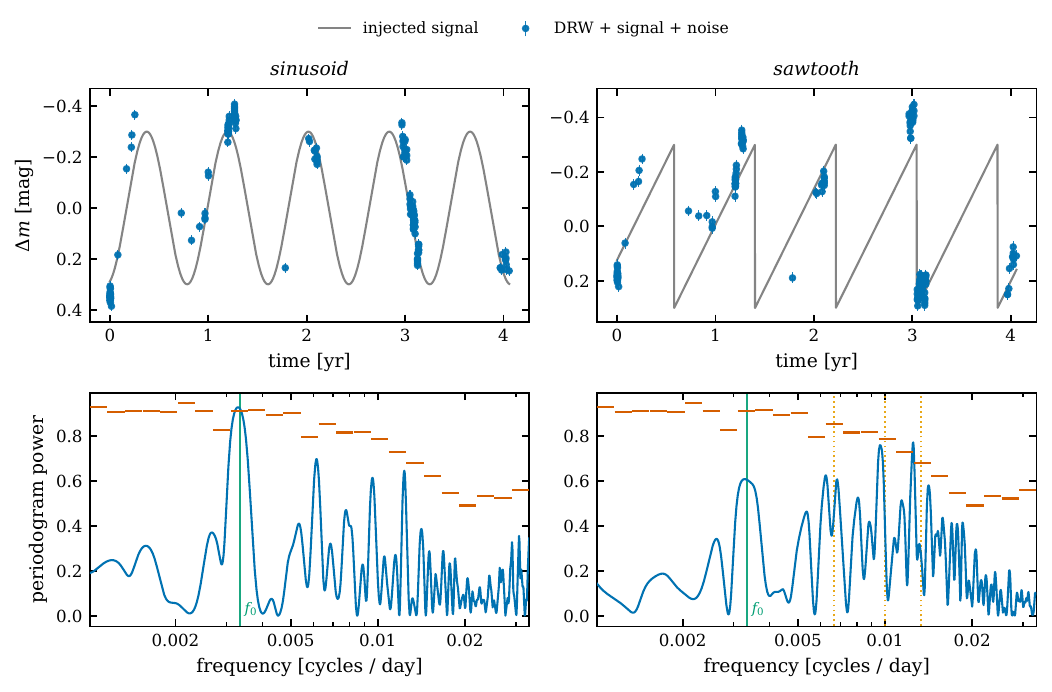}
\caption{Illustration of the detection test for one system on the median-baseline PTF cadence of Figure~\ref{fig:cadences}, with $\log_{10}(M_{\rm BH}/M_\odot)=8.5$, $M_i=-24$, $z=1$, $P=300$\,d and $A=0.3$\,mag. \textit{Top:} injected signal (grey) and simulated light curve (dots) for the sinusoid (left) and sawtooth (right). \textit{Bottom:} periodogram (blue), per-bin thresholds from $N_{\rm MC}=2000$ null light curves (orange), injected frequency $f_0$ (green) and, for the sawtooth, its harmonics (dotted).}
\label{fig:detection_anatomy}
\end{figure}

For each system we draw $N_{\rm MC}$ null light curves (MC for Monte Carlo) at the same cadence and DRW parameters. A null light curve is a noise-only light curve with no injected signal, a DRW realisation with the system's $\tau$ and $\sigma$ plus Gaussian photometric noise at the per-epoch error of the cadence. With periodograms $p_j(f)$ for the null light curves, $j=1,\ldots,N_{\rm MC}$, the per-bin maximum power across them is the threshold, and an injected light curve with periodogram $p(f)$ is significant when it clears that threshold,
\begin{equation}
\max_{f\in B_k} p(f) \;>\; \max_{j}\,\max_{f\in B_k} p_j(f) ,
\label{eq:detect}
\end{equation}
in at least one of the 25 bins $B_k$ (Figure~\ref{fig:detection_anatomy}), the LCH26 rule, mirroring their maximum over $10^5$ null light curves. One set of null light curves, and so one set of 25 thresholds, serves both signal shapes of a system.

We call all the injections on one type of cadence, with one rule for the magnitudes and one $N_{\rm MC}$, a run. The main PTF, Stripe~82 and LSST-like runs, with $M_i$ drawn from DR7, use $N_{\rm MC}=10^5$, matched to LCH26, and we call them the DR7-magnitude runs. A threshold set by the maximum over $N_{\rm MC}$ null light curves keeps creeping up with $N_{\rm MC}$, so $10^5$ makes the comparison like for like without making the absolute rates converged (Section~\ref{sec:limitations} measures by how much). A few checks use fewer null light curves and carry a dagger.\footnote{Numbers from runs below $N_{\rm MC}=10^5$, and quantities derived from them such as differences and their significances, are marked with a dagger ($^{\dagger}$). A run with fewer null light curves per system has lower thresholds, so its fractions are compared only with values from runs at the same $N_{\rm MC}$, except where Section~\ref{sec:limitations} measures the effect of $N_{\rm MC}$ itself and where Figure~\ref{fig:daily} sets the LCH26 values beside them.}

The $N_{\rm MC}+1$ light curves of a system are exchangeable under the null hypothesis (drawn from one distribution, so every ordering of them is equally likely), and that caps the probability that a null light curve is counted significant: it takes the strict record in a given bin with probability at most $1/(N_{\rm MC}+1)$, so it clears at least one of the 25 bins with probability at most $25/(N_{\rm MC}+1)=2.5\times10^{-4}$ at $N_{\rm MC}=10^5$. Appendix~\ref{app:fap} measures where under that cap the rule sits for the three DR7-magnitude runs.

\subsection{Recovery definitions}
\label{sec:definitions}

We carry two recovery definitions throughout, because the two studies have to be compared on one criterion. Significance-only recovery counts any injection whose light curve clears the threshold in some bin (Equation~\eqref{eq:detect}), and it is the direct match to their main recovery table. Significance + period recovery adds the requirement (the period check) that the recovered peak fall within 15 per cent of the injected period,
\begin{equation}
|P_{\rm peak}-P|/P<0.15 ,
\label{eq:pcheck}
\end{equation}
where the recovered peak, at period $P_{\rm peak}$, is the tallest point of the periodogram over the whole grid, which need not lie in the bin that cleared the threshold. In the example of Figure~\ref{fig:detection_anatomy} (thresholds from $N_{\rm MC}=2000$), the sinusoid periodogram peaks once at 303.9\,d, inside the 15 per cent window, and clears the threshold in one bin, so it counts on both definitions. The sawtooth spreads its power across the fundamental and the harmonics above it, and its tallest peak lands at 79.9\,d, between the third and fourth harmonics, clearing the threshold in two bins while missing the injected period. LCH26 report the period-match fraction separately: on their PTF-like cadence, 97.0 per cent of significant sinusoid detections and 79.6 per cent of sawtooth detections land within that window, and Table~\ref{tab:ptf} gives the significance + period rates these shares imply.

\subsection{Sampling across mass and error estimates}
\label{sec:sampling}

To span the mass range we split the catalogue into eight 0.5-dex bins over $\log M_1\in[6,10)$ and draw a fixed count per bin, 375 systems in each without replacement, so each DR7-magnitude run holds 3000 systems out of the 18,374 and 1500 of them land in the high-mass range. The Stripe~82 and PTF runs use the same 3000 systems, drawn in two halves ($\log M_1$ from 6 to 8 and from 8 to 10), and the LSST-like run is a separate draw of 3000 that shares 1853 of them. In the PTF run each system is assigned one of the 49 cadences by a fixed function of its row in the merger catalogue, effectively uniform with replacement, so each cadence is assigned to 44 to 80 of the 3000 systems. The draw of the systems, the draw of the magnitudes, the noise and injections of each system and the assignment of the PTF cadences use fixed seeds, which Appendix~\ref{app:draws} gives with the cadence-assignment rule.

We quote two averages over the high-mass range. The average with equal mass weight is the plain mean over its 1500 systems, 375 per bin, and averages over the full range $\log M_1\in[6,10)$ use equal mass weight too. The average with DR7 mass weight reweights the same high-mass systems to the \citet{Shen2011} DR7 black hole mass distribution, that of the 99,409 DR7 quasars with $\log_{10}(M_{\rm BH}/M_\odot)\in[8,10)$ in 20 bins of 0.1\,dex, with weight $p_{\rm DR7}(b)/p_{\rm sample}(b)$ for bin $b$, normalised to a mean of one, and its errors use the effective sample size of the weights, $(\sum w)^2/\sum w^2$ (Table~\ref{tab:ptf} gives both averages). Errors are binomial, except on PTF, where systems sharing a cadence are correlated and the binomial error is multiplied by the square root of the design effect, the factor by which that correlation inflates the variance,
\begin{equation}
D = 1+(\bar m-1)\rho ,
\label{eq:deff}
\end{equation}
where $\bar m$ is the mean number of systems per cadence (30.6 for the 1500 high-mass systems) and $\rho$ the intraclass correlation, the correlation between outcomes of systems on the same cadence, estimated by a one-way analysis of variance (from the variance of the outcomes between cadences and within them). The comparisons of Section~\ref{sec:daily} and Appendix~\ref{app:nights} use cluster-robust errors, which treat the systems on one cadence as one group: the sandwich estimator, which takes the standard error from the scatter of the per-cadence sums of residuals, with the 49 cadences as clusters and no correction for their small number. Differences and ratios quoted in the text are computed from unrounded values and rounded once, so a recomputation from printed operands can disagree in the last digit.

Four further runs, each with the period range and errors of the DR7-magnitude run on the same cadence, change how the magnitude is assigned, $N_{\rm MC}$ or both. The LSST-like run with $M_i=-23$ for every system, at $N_{\rm MC}=10^5$, holds 250 systems in each of the seven lower mass bins and 175 in the top bin, 1925 in all (Table~\ref{tab:lsst}). At $N_{\rm MC}=10^4$, a Stripe~82 run with $M_i=-23$ uses 200 systems per bin, a Stripe~82 run with $M_i$ drawn from DR7 100 per bin, and a PTF run with the same cadences and magnitudes 250 per bin.

\section{Results: PTF and Stripe 82 Cadences}
\label{sec:results-ptf-s82}

An archival search that adopts the LCH26 fractions as its detection efficiency rests on two premises: that recovery does not change with black hole mass, and that an independent implementation of the method reproduces the published PTF values. We tested the first by measuring recovery in each mass bin on the observed PTF and Stripe~82 cadences, and the second by comparing the high-mass PTF values with LCH26. Table~\ref{tab:ptf} sets the PTF run against LCH26 in both definitions and both weightings, with Stripe~82 beside it, and Table~\ref{tab:lsst} gives the LSST-like run taken up in Section~\ref{sec:results-lsst}.

\begin{figure}[tbp]
\centering
\includegraphics[width=\textwidth]{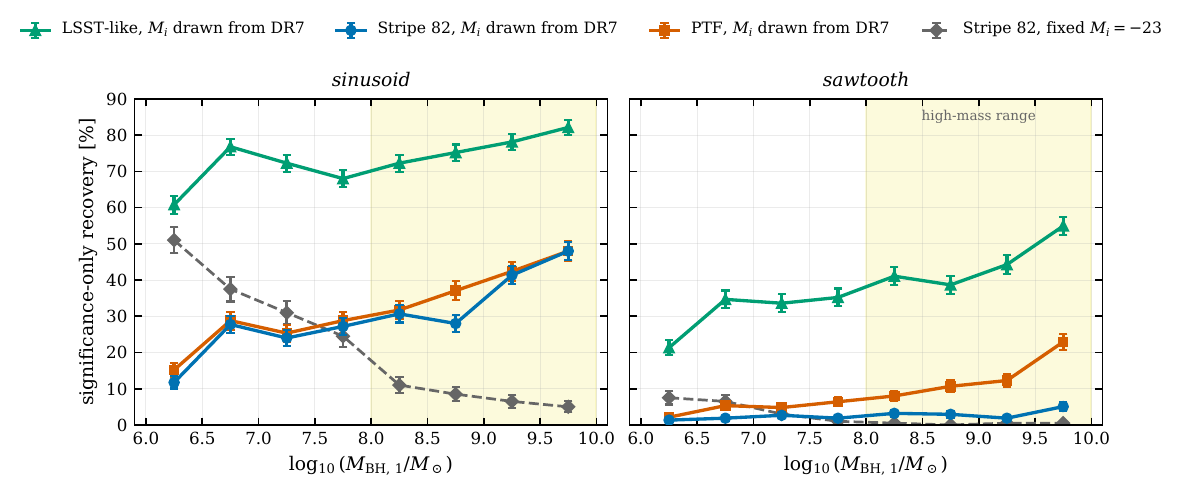}
\caption{Significance-only recovery (no period check) against primary black hole mass in eight 0.5-dex bins, for sinusoids (left) and sawtooth signals (right), with $M_i$ drawn from DR7 and $N_{\rm MC}=10^5$ on the LSST-like (triangles), Stripe~82 (circles) and PTF (squares) cadences, and with $M_i=-23$ and $N_{\rm MC}=10^4$ on the Stripe~82 cadence (diamonds, dashed; only its trend compares). Periods are drawn from $U[30,779]$\,d, and from $U[200,1100]$\,d on Stripe~82. The shaded band is the high-mass range, $\log M_1\in[8,10)$. Errors are binomial, widened on PTF for the shared cadences.}
\label{fig:recovery_mass}
\end{figure}

\begin{table}[htbp]
\centering
\caption{Recovery fractions (per cent) on the PTF and Stripe~82 cadences for $\log_{10}(M_{\rm BH,1}/M_\odot)\in[8,10)$, compared with LCH26 \citep{Lin2026}.}
\label{tab:ptf}\label{tab:s82}
\begin{tabular*}{\textwidth}{@{\extracolsep{\fill}}llccccc@{}}
\toprule
 & & \multicolumn{3}{c}{PTF} & \multicolumn{2}{c}{Stripe~82} \\
\cmidrule(lr){3-5}\cmidrule(lr){6-7}
 & & Equal mass & DR7 mass & & Equal mass & DR7 mass \\
Signal & Definition & weight & weight & LCH26 & weight & weight \\
\midrule
sinusoid & significance-only & $39.8\pm1.6$ & $40.2\pm1.8$ & 44.7 & $37.0\pm1.2$ & $36.4\pm1.4$ \\
         & significance + period & $37.9\pm1.7$ & $38.2\pm1.9$ & 43.4 & $\cdots$ & $\cdots$ \\
sawtooth & significance-only & $13.5\pm0.9$ & $12.6\pm1.0$ & 9.3 & $3.3\pm0.5$ & $3.0\pm0.5$ \\
         & significance + period & $9.0\pm0.8$ & $8.5\pm0.9$ & 7.4 & $\cdots$ & $\cdots$ \\
\bottomrule
\end{tabular*}
\par\leftskip=0pt\rightskip=0pt\parfillskip=0pt plus 1fil\relax
\tablenotes{Equal mass weight is the mean over the 1500 systems per signal shape in this mass range, 375 in each 0.5-dex bin. DR7 mass weight reweights the same systems to the black hole mass distribution of the \citet{Shen2011} DR7 quasars. PTF errors are binomial errors widened by $\sqrt{1+(\bar m-1)\rho}$, where $\bar m=30.6$ systems share each of the 49 PTF cadences and $\rho$ is the one-way analysis-of-variance correlation between systems on the same cadence, measured for each row ($0.0173$, $0.0268$, $-0.0061$ and $0.0044$ in row order; a negative value is set to zero). The LCH26 significance + period values are their significance-only rates multiplied by the share of their significant detections whose highest peak lies within 15 per cent of the injected period, 0.970 for the sinusoid and 0.796 for the sawtooth. Stripe~82 errors are binomial, and only significance-only recovery is given there; its injected periods are drawn from $U[200,1100]$\,d, against $U[30,779]$\,d on PTF, so no LCH26 case matches the Stripe~82 run.}
\end{table}

\begin{table}[htbp]
\centering
\caption{Recovery fractions (per cent) on the LSST-like cadence for $\log_{10}(M_{\rm BH,1}/M_\odot)\in[8,10)$, compared with LCH26 \citep{Lin2026}.}
\label{tab:lsst}
\begin{tabular*}{\textwidth}{@{\extracolsep{\fill}}lcc@{}}
\toprule
Source & Sinusoid & Sawtooth \\
\midrule
This work, $M_i$ drawn from DR7, significance-only & $76.9\pm1.1$ & $44.7\pm1.3$ \\
This work, $M_i$ drawn from DR7, significance + period & $75.5\pm1.1$ & $42.5\pm1.3$ \\
This work, fixed $M_i=-23$, significance-only & $49.5\pm1.6$ & $22.4\pm1.4$ \\
This work, fixed $M_i=-23$, significance + period & $46.6\pm1.6$ & $16.9\pm1.2$ \\
LCH26, significance-only & 22.6 & 1.1 \\
LCH26, significance + period & 22.5 & 0.97 \\
\bottomrule
\end{tabular*}
\par\leftskip=0pt\rightskip=0pt\parfillskip=0pt plus 1fil\relax
\tablenotes{Both runs of this work use the 484-epoch LSST-like cadence and $N_{\rm MC}=10^5$ null light curves per system. The two runs differ in how $M_i$ is set and in which systems they draw: 1500 systems per signal shape with $M_i$ drawn from DR7, 925 with $M_i=-23$. Errors are binomial. The fixed-$M_i$ run holds 250, 250, 250 and 175 systems in the four 0.5-dex bins; with equal bin weights its sinusoid value is 48.9 per cent. The LCH26 significance + period values are their significance-only rates multiplied by the share of their significant detections whose highest peak lies within 15 per cent of the injected period, 99.7 per cent for the sinusoid and 87.8 per cent for the sawtooth (their \S3.1).}
\end{table}

\begin{table}[htbp]
\centering
\caption{Significance-only recovery fractions (per cent) in each 0.5-dex bin of primary black hole mass, with $M_i$ drawn from DR7 and $N_{\rm MC}=10^5$.}
\label{tab:perbin}
\begin{tabular*}{\textwidth}{@{\extracolsep{\fill}}lcccccc@{}}
\toprule
 & \multicolumn{3}{c}{Sinusoid} & \multicolumn{3}{c}{Sawtooth} \\
\cmidrule(lr){2-4}\cmidrule(lr){5-7}
$\log_{10}(M_{\rm BH,1}/M_\odot)$ & Stripe~82 & PTF & LSST-like & Stripe~82 & PTF & LSST-like \\
\midrule
$[6.0,6.5)$ & $11.7\pm1.7$ & $15.2\pm2.0$ & $60.8\pm2.5$ & $1.3\pm0.6$ & $2.1\pm0.7$ & $21.3\pm2.1$ \\
$[6.5,7.0)$ & $27.7\pm2.3$ & $28.8\pm2.5$ & $76.8\pm2.2$ & $1.9\pm0.7$ & $5.3\pm1.2$ & $34.7\pm2.5$ \\
$[7.0,7.5)$ & $24.0\pm2.2$ & $25.3\pm2.4$ & $72.3\pm2.3$ & $2.7\pm0.8$ & $4.8\pm1.1$ & $33.6\pm2.4$ \\
$[7.5,8.0)$ & $27.2\pm2.3$ & $28.8\pm2.5$ & $68.0\pm2.4$ & $1.9\pm0.7$ & $6.4\pm1.3$ & $35.2\pm2.5$ \\
$[8.0,8.5)$ & $30.7\pm2.4$ & $31.7\pm2.5$ & $72.3\pm2.3$ & $3.2\pm0.9$ & $8.0\pm1.4$ & $41.1\pm2.5$ \\
$[8.5,9.0)$ & $28.0\pm2.3$ & $37.1\pm2.6$ & $75.2\pm2.2$ & $2.9\pm0.9$ & $10.7\pm1.6$ & $38.7\pm2.5$ \\
$[9.0,9.5)$ & $41.3\pm2.5$ & $42.4\pm2.7$ & $78.1\pm2.1$ & $1.9\pm0.7$ & $12.3\pm1.7$ & $44.3\pm2.6$ \\
$[9.5,10.0)$ & $48.0\pm2.6$ & $48.0\pm2.7$ & $82.1\pm2.0$ & $5.1\pm1.1$ & $22.9\pm2.2$ & $54.9\pm2.6$ \\
\bottomrule
\end{tabular*}
\par\leftskip=0pt\rightskip=0pt\parfillskip=0pt plus 1fil\relax
\tablenotes{Each entry is the fraction of the 375 systems in that bin whose injected signal is significant, averaged over the redshifts of those systems, over amplitudes $A\sim U[0,0.5)$\,mag and at the per-epoch error of the cadence (0.02, 0.05 and 0.01\,mag on Stripe~82, PTF and LSST-like). Errors are binomial; on PTF they are widened by the square root of the per-bin design effect from the 49 shared cadences, as in Figure~\ref{fig:recovery_mass}. Injected periods are drawn from $U[30,779]$\,d on PTF and LSST-like and from $U[200,1100]$\,d on Stripe~82. Magnitudes in the lowest bin are extrapolated below the DR7 mass range, and the second bin and the lower part of the third draw on the same DR7 quasars. DR7 is flux limited, so the magnitudes drawn at low mass are brighter than typical for that mass (Section~\ref{sec:mass-range}).}
\end{table}

\subsection{Recovery against black hole mass}
\label{sec:mass}

LCH26 give each recovery fraction for their whole sample, and a forecast that applies it across a mass function is right only if recovery does not change with mass. We measured significance-only recovery (no period check) in each mass bin on the Stripe~82 and PTF cadences with $M_i$ drawn from DR7, and on the Stripe~82 and LSST-like cadences with $M_i=-23$ for every system. At a fixed $M_i=-23$ the DRW amplitude rises with mass, and the Stripe~82 sinusoid recovery falls from \prov{51.0} per cent in the lowest mass bin to \prov{5.0} per cent in the highest (Figure~\ref{fig:recovery_mass}). The LSST-like cadence shows the same fall at $N_{\rm MC}=10^5$, from 78.0 per cent in the lowest bin to 41.1 per cent in the highest for the sinusoid and from 44.4 to 22.9 per cent for the sawtooth. Under the DR7 draw the high-mass systems brighten. Across the eight 0.5-dex bins the PTF sinusoid climbs from 15 to 48 per cent and the Stripe~82 sinusoid from 11.7 to 48.0 per cent (Table~\ref{tab:perbin}), while the Stripe~82 sawtooth stays between 1.3 and 5.1 per cent at every mass.

Which way recovery runs with mass follows from the magnitude draw. The DR7 relation has slope $dM_i/d\log M_{\rm BH}=-1.81$, so heavier black holes are brighter (Figure~\ref{fig:mi_mechanism}). Fed through Equation~\eqref{eq:sigma}, whose mass coefficient is 0.18 and whose luminosity term is $0.131\,(M_i+23)$, the net slope of DRW amplitude with mass under the drawn magnitudes is
\begin{equation}
\frac{d\log\sigma}{d\log M_{\rm BH}} \approx 0.18 + 0.131\times(-1.81)
   \approx -0.06,
\label{eq:slope}
\end{equation}
close to flat, against $+0.18$ at fixed $M_i$. Recovery falls as the noise amplitude rises, so the sign of the mass trend is set by how the magnitude is assigned. The $-1.81$ is an ordinary least-squares fit over the whole catalogue, and the relation underneath it is curved. Between $\log_{10}(M_{\rm BH}/M_\odot)=6.502$ and about 7.3 the median magnitude of the draw stops moving with mass, because every mass there draws from the same 200 lowest-mass DR7 objects (Section~\ref{sec:magnitudes}), and the local net amplitude slope is $+0.18$, the fixed-$M_i$ value. The $-0.06$ applies above $\log M_1\approx8$ and, through the extrapolated relation, below 6.502, and in between the local slope passes from one value to the other.

\subsection{Comparison with published PTF recovery}
\label{sec:ptf-lch26}

The LCH26 PTF fractions (Table~\ref{tab:ptf}) are the published Lomb-Scargle recovery fractions for PTF light curves, and so the detection efficiencies that a search of the PTF archive would adopt.

\begin{figure}[tbp]
\centering
\includegraphics[width=\textwidth]{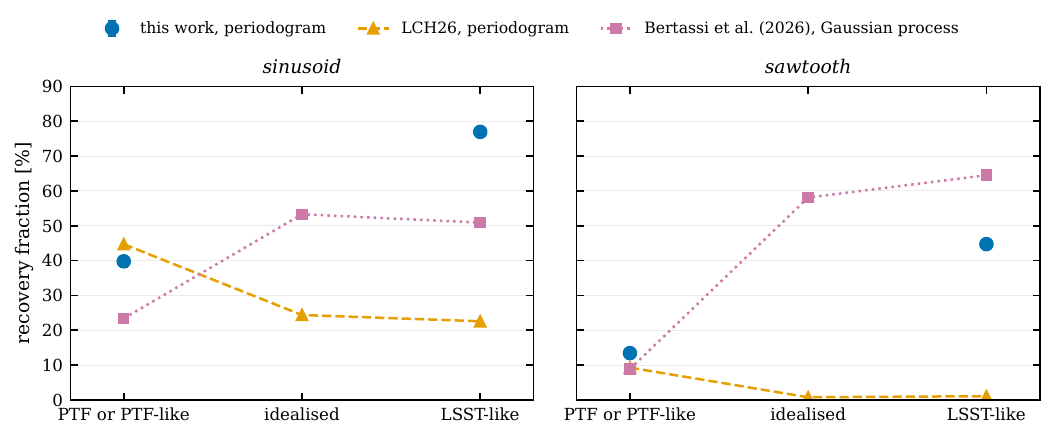}
\caption{Significance-only recovery (no period check) by cadence for sinusoids (left) and sawtooth signals (right) in three studies: this work (circles; PTF and LSST-like only, $\log M_1\in[8,10)$, $M_i$ drawn from DR7, $N_{\rm MC}=10^5$; daily sampling is in Figure~\ref{fig:daily}), the LCH26 periodogram (triangles) and the Gaussian-process model of \citet{Bertassi2026} (squares; Bayes-factor thresholds $\log_{10}B=5$, or 3 at LSST-like cadence; a different method, so only the order of its cadences compares), both for the full LCH26 sample, whose idealised cadence samples the PTF-like baselines daily. Our errors are binomial, widened on PTF.}
\label{fig:vs_lch26}
\end{figure}

In the high-mass range with equal mass weight, the PTF run recovers $39.8\pm1.6$ per cent of sinusoids on significance-only, against 44.7 per cent in LCH26 (Table~\ref{tab:ptf}, Figure~\ref{fig:vs_lch26}). The difference is 4.9 percentage points (hereafter points), 3.9 binomial standard errors for 1500 systems. The 49 cadences are shared across systems, and the measured intraclass correlation of 0.017 gives a design effect of 1.51 (Equation~\eqref{eq:deff}), so the standard error widened for the shared cadences is 1.55 points against the binomial 1.26 and the gap reads 3.1 standard errors. The noise generator sits under this comparison as a further construction difference. LCH26 build their DRW light curves the \citet{TimmerKoenig1995} way, daily over the observed baseline, which we read as a generator without padding (padding means drawing a longer series and keeping a stretch of the light curve's length), where we use the Kelly recursion of Equation~\eqref{eq:ou}. On one system at 0.02\,mag errors and no redshift (Appendix~\ref{app:generator}) the unpadded generator raises significance-only PTF recovery by \prov{$19.2\pm1.3$} points over the recursion, the direction in which their value exceeds ours, while the padded one agrees with it, so if LCH26 did not pad, the generator is a candidate for part of the gap, by an amount one system cannot fix for their sample.

Part of the 4.9-point difference has a clean and one-sided cause. We draw the injected periods from one range for every cadence while the search grid of each light curve starts at $f_{\rm min}=1.5/T$, and 22 of the 49 cadences are short enough that $T/1.5<779$\,d, so some injections sit where the grid cannot reach the injected frequency at all. Restricting the run to the injections with $P\le T/1.5$, which is the LCH26 construction, closes 1.4 of the 4.9-point sinusoid difference and widens the sawtooth excess (Appendix~\ref{app:ptfdiff}). The remaining 3.5 points have no single identified cause, and Appendix~\ref{app:ptfdiff} lists the candidates.

The sawtooth runs the other way: on significance-only the PTF run returns $13.5\pm0.9$ per cent against the LCH26 9.3. A DRW-amplitude error would move both signal shapes together, and here the sinusoid sits below LCH26 and the sawtooth above, so a shared amplitude bias is not what lifts the sawtooth. Harmonic leakage covers about 60 per cent of it: a sawtooth deposits power at the fundamental and its harmonics, and 66.8 per cent of our significant sawtooth detections peak within 15 per cent of the injected period against 79.6 per cent in LCH26 (Figure~\ref{fig:detection_anatomy}), which accounts for 2.6 of the 4.2 points by which our sawtooth recovery exceeds theirs (Appendix~\ref{app:ptfdiff}).

\subsection{Recovery on the Stripe 82 cadence}
\label{sec:s82}

LCH26 give no recovery for a record as long and sparse as a Stripe~82 light curve, which is the efficiency a search of the Stripe~82 light curves of DR7 quasars would need. The Stripe~82 cadence (Section~\ref{sec:cadences}) has only 60 nights over nine years with deep seasonal gaps, and it still recovers $37.0\pm1.2$ per cent of sinusoids and $3.3\pm0.5$ per cent of sawtooth signals on significance-only in the high-mass range with equal mass weight, holding at $36.4\pm1.4$ and $3.0\pm0.5$ per cent with DR7 mass weight (Table~\ref{tab:ptf}). The Stripe~82 run uses the $U[200,1100]$\,d period range on a cadence with no LCH26 counterpart, so it stands as an archival measurement in its own right. The period range and the 0.02\,mag errors also differ from the PTF run, so the gap between the Stripe~82 and PTF curves of Figure~\ref{fig:recovery_mass} is not a clean cadence comparison.

\section{Results: LSST-like Cadence}
\label{sec:results-lsst}

The number of binaries an LSST forecast predicts scales with the LSST-like recovery fraction, so this section asks whether an independent implementation of the method reproduces the LCH26 value and which inputs set it. Table~\ref{tab:lsst} sets the LSST-like run against the LCH26 LSST-like value, and Table~\ref{tab:inputs} lists each input in which our setup differs from theirs or their text allows more than one reading, with its test and its effect.

\subsection{Comparison with published LSST-like recovery}
\label{sec:lsst-lch26}

If the method and the inputs LCH26 state fix their LSST-like fractions, an independent implementation returns them, and any difference passes directly into an LSST forecast. Both runs of Table~\ref{tab:lsst} use the 484-epoch LSST-like cadence at $N_{\rm MC}=10^5$, and the $M_i$ assignment is the only setting that changes between them, though the two runs also draw on different systems. With $M_i$ drawn per system from DR7, the sinusoid recovers $76.9\pm1.1$ per cent on significance-only in the high-mass range, against 22.6 per cent in LCH26, a ratio of about 3.4, and the ratio is also 3.4 with the period check (Figure~\ref{fig:vs_lch26}). The sawtooth recovers $44.7\pm1.3$ per cent against their 1.1 per cent, a ratio of 40.7, and the sawtooth ratios, with and without the period check, are the widest in the whole comparison. Reweighting the high-mass systems to the DR7 mass histogram lands the sinusoid at $77.0\pm1.2$ per cent and the sawtooth at $43.4\pm1.5$ per cent, a shift of $-1.4$ points for the sawtooth (from unrounded values). Over the full $\log M_1\in[6,10)$ range the sinusoid holds $73.2\pm0.8$ per cent and the sawtooth $38.0\pm0.9$ per cent, so the high recovery is not confined to the massive end. Within the high-mass range the sinusoid climbs from 72.3 per cent in $[8.0,8.5)$ to 82.1 per cent in $[9.5,10.0)$, a rise of 3.2 standard errors, and that climb is the part of the shape worth reading (Table~\ref{tab:perbin}, Figure~\ref{fig:recovery_mass}). The low end is not monotone, though at 375 systems per bin the structure there is weak: the third bin lies below the second (72.3 against 76.8 per cent), as the magnitude draw predicts, since the two bins share magnitudes and the third sits at higher mass, where the fixed-$M_i$ slope of $+0.18$ applies (Section~\ref{sec:mass}). Read at the DRW amplitude scale of their Figure~3, the sinusoid ratio is at least 2.7 (Appendix~\ref{sec:amplitude}). Among the inputs tested in this section we have not identified the source of most of the difference, and the list of injected periods or the detection code of LCH26 would allow a direct comparison.

The magnitude assignment is worth isolating, because the same cadence was also run with $M_i=-23$ held fixed for every system, at the same $N_{\rm MC}$. The fixed-$M_i$ run returns 49.5 per cent on the sinusoid and 22.4 per cent on the sawtooth (Table~\ref{tab:lsst}), so the DR7 draw raises the sinusoid by 27.4 points and the sawtooth by 22.4 points. Even at fixed $M_i$ the recovery is 2.2 times their value for the sinusoid and 20 times for the sawtooth, so the difference does not come from magnitudes that are too bright on our side. Inside the high-mass range the drawn magnitudes are bright, median $M_i=-25.96$ with 5th and 95th percentiles at $-27.90$ and $-23.20$, and 96.9 per cent of these systems sit brighter than the fixed $-23$. The DRW amplitude falls with luminosity through the $0.131\,(M_i+23)$ term, and a quieter light curve leaves the injected signal more room. The high-mass range holds 925 systems in the fixed-$M_i$ run against 1500 in the other, so what the pair holds fixed is the cadence and $N_{\rm MC}$ and not the systems.

\begin{figure}[tbp]
\centering
\includegraphics[width=\textwidth]{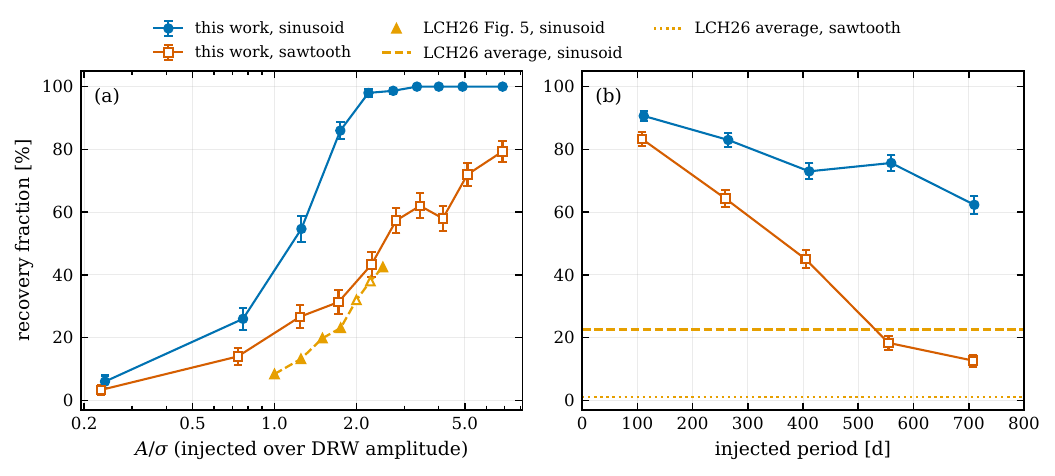}
\caption{Significance-only recovery (no period check) at LSST-like cadence for $\log M_1\in[8,10)$, with $M_i$ drawn from DR7 and $N_{\rm MC}=10^5$, for sinusoids (circles) and sawtooth signals (squares). (a) Recovery against $A/\sigma$, injected amplitude over DRW amplitude, in ten bins of 150 injections, with the LCH26 LSST-like sinusoid curve read from their Figure~5 (triangles, open where it merges with their idealised curve). (b) Recovery against injected period in five bins of 300 injections, with the LCH26 LSST-like averages over their own periods (horizontal lines). Errors are binomial.}
\label{fig:lsst_amp_period}
\end{figure}

\subsection{Recovery at the same mean $A/\sigma$}
\label{sec:snr}

Recovery rises with $A/\sigma$, the ratio of the injected amplitude to the DRW amplitude, so if the two studies inject at different $A/\sigma$, each fraction holds only for its own amplitude distribution and a forecast could not carry either over to a survey. A shorter reading of the LSST-like difference is that our injected signals sit louder against their own noise, and conditioning both sides on that ratio tests it directly. The LCH26 Figure~5 plots recovery against $A/\sigma$ for the LSST-like cadence, and reading their dotted LSST-like curve at the mean $A/\sigma$ of our high-mass systems in three bins from 1.0 to 2.5 (Figure~\ref{fig:lsst_amp_period}a shows the curve and their digitised values) leaves the difference standing: we return 53.2, 84.7 and 98.7 per cent where their curve reads 12.8, 23.1 and 37.8 per cent, so 4.2, 3.7 and 2.6 times their value bin by bin. The top bin reads their curve where its LSST-like and idealised-cadence branches merge, and reading the merged curve can only lift their value and shrink the ratio. We read their $A/\sigma$ axis with $\sigma$ as in Equation~\eqref{eq:sigma}; if it is $A/\mathrm{SF}_\infty$, their curve sits lower and every ratio rises. Above $A=0.45$\,mag our high-mass LSST-like run recovers 142 of 142 injected sinusoids, and the published LCH26 curves of recovery against injected amplitude top out near 80 per cent at $A=0.5$\,mag. The ceiling is the sharpest quantitative tension in the whole comparison, and no third source settles it, since no amplitude-resolved periodogram recovery curve at LSST-like cadence, with a threshold in each frequency bin, has been published other than their Figure~5. The amplitude of what we inject is therefore not what separates the two studies.

\subsection{Recovery with one epoch per day over the PTF baselines}
\label{sec:daily}

Whether denser sampling lowers recovery decides whether a forecast for a survey that samples more densely than PTF should expect fewer detections, as LCH26 conclude. LCH26 run three cadences, and they do not order by baseline: PTF-like at 44.7 per cent over 3.2\,yr on about 98 nights, a count which their 12-day mean spacing implies and which they never state, idealised at 24.4 per cent on daily sampling of those same PTF baselines, which has more than ten times the epochs, and LSST-like on 5-day sampling over 10\,yr. The LCH26 split makes the point: the fall from 44.7 to 24.4 per cent happens at one baseline, while the further step to their LSST-like value is under two points as the baseline goes from 3.2 to 10\,yr and the photometric error from 0.05 to 0.01\,mag. The axis they identify is therefore sampling density. LCH26 say as much in their \S4: in the sparser PTF-like light curves ``the sparse sampling can in many cases be advantageous'', because the observed epochs can by chance land near the maxima and minima of an injected signal, and LCH26 flag their own pattern as ``a surprising (to us at least) finding''.

\begin{figure}[tbp]
\centering
\includegraphics[width=\textwidth]{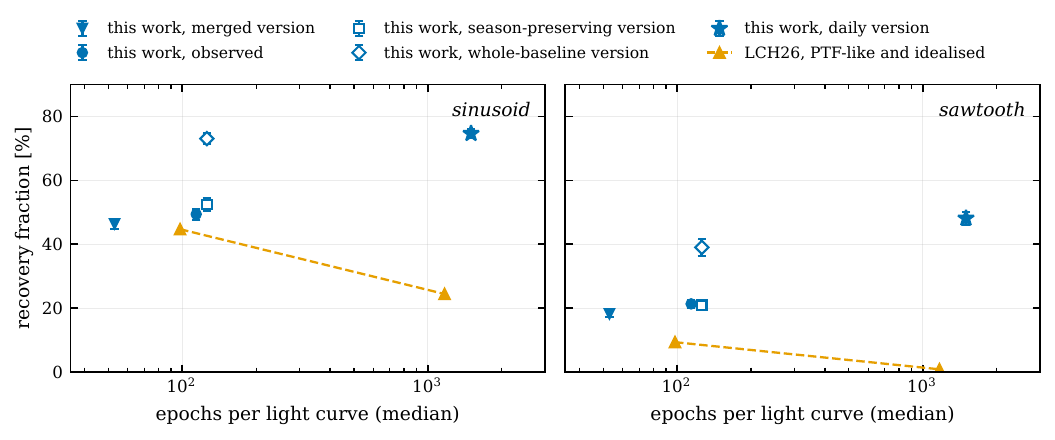}
\caption{Significance-only recovery (no period check) for $\log M_1\in[8,10)$ on the 49 PTF baselines at $N_{\rm MC}=10^4$, against median epochs per light curve, for sinusoids (left) and sawtooth signals (right): observed cadences with and without repeat exposures merged, two rebuilds at the LCH26 mean night spacing (open markers) and daily sampling. Triangles are the LCH26 PTF-like and idealised values ($N_{\rm MC}=10^5$; only the direction of each step compares), at the epoch counts their text implies. Errors are clustered over the 49 cadences.}
\label{fig:daily}
\end{figure}

We ran that construction. The daily version of each of our 49 PTF cadences has one epoch per day from the first observed epoch to within half a day of the last, $N={\rm round}(T)+1$ epochs (744 to 2034, median 1483, against a median of 114 observed). The run with daily sampling keeps the 0.05\,mag photometric error and the $U[30,779]$\,d injected periods of our PTF run, and assigns the cadences exactly as the PTF run does, with 375 systems per bin. The run used $N_{\rm MC}=10^4$, coarser than the main runs and below the $10^5$ LCH26 use, so every number from it is set only against runs at the same $10^4$, and their 24.4 per cent is not the comparator for our daily value (Figure~\ref{fig:daily}). In the high-mass range the sinusoid clears the threshold in \prov{$74.7\pm1.3$} per cent of injections with daily sampling, where the observed cadences return \prov{$49.4\pm1.7$} per cent, and the sawtooth in \prov{$48.1\pm2.1$} against \prov{$21.3\pm1.3$} per cent; errors on these recovery fractions are cluster-robust on the 49 shared cadences (Section~\ref{sec:sampling}). Daily sampling of the same baselines therefore raises recovery by \prov{25.3} points for the sinusoid (\prov{11.6} standard errors) and \prov{26.8} points for the sawtooth. Each rise is the difference of two recovery fractions. The other estimator, the mean of the 49 per-cadence differences, gives \prov{$25.4\pm2.2$} points (\prov{11.8} standard errors), with sinusoid recovery rising on \prov{47} of the 49 cadences. On the same comparison LCH26 find that recovery falls, for the sawtooth from 9.3 to 0.8 per cent.

The night structure of the PTF records is a further unmatched input. Our PTF cadences average 57 distinct nights per cadence against about 98 implied by their stated mean spacing, and the daily version escapes this input entirely, since a daily grid misses no night. Fewer nights depress recovery on the observed cadences, so the difference could inflate the rise measured against them. Appendix~\ref{app:nights} measures it with three rebuilt versions of the 49 cadences: a merged version, with the repeat exposures of each night merged into one epoch, and two versions at the LCH26 mean spacing of about 12\,d between nights, a season-preserving version with the nights placed at random inside the observing seasons of each cadence (stretches of nights with no gap longer than 60\,d) and a whole-baseline version with them placed at random anywhere in its baseline (Table~\ref{tab:inputs} gives the results). What their text does not settle is which placement of the nights is theirs, since one published sampling statistic, a mean night spacing, does not discriminate between the two. Against every version, daily sampling raises sinusoid recovery or leaves it unchanged within its error. The test runs our pipeline on their construction, so it rules sampling density out as the axis on our side, and it does not say what their pipeline does differently.

\subsection{Distribution of injected periods}
\label{sec:periods}

Recovery at LSST-like cadence falls with injected period (Figure~\ref{fig:lsst_amp_period}b), so the period distribution, which differs between the two studies, could by itself change the LSST-like fraction a forecast takes from either. The LCH26 footnote~3 hands the LSST-like set the periods already drawn for the PTF-like set, so what faces our $U[30,779]$\,d range (Section~\ref{sec:signals}) is a mixture of per-light-curve $U[0,T/1.5]$ draws, one uniform per PTF baseline. The mixture puts mass below our 30\,d lower limit and, for the longer PTF baselines, above our 779\,d upper limit. We have not run their period distribution end to end, so no single corrected number is quoted for it. The bounds of Appendix~\ref{app:periods} are enough to settle what it can do: they put our LSST-like recovery under their mixture of periods above their values for both signal shapes, so the period distribution is bounded and cannot carry the difference.

\begin{table}[tbp]
\centering
\caption{Inputs of the comparison with LCH26 that differ from theirs or that
their text leaves open, and the test of each.}
\label{tab:inputs}
\setlength{\tabcolsep}{2pt}
\begin{tabular*}{\textwidth}{@{\extracolsep{\fill}}lllll@{}}
\toprule
Input & This work & LCH26 & Effect & Section \\
\midrule
\parbox[t]{2.2cm}{\raggedright Injected periods} &
  \parbox[t]{3.3cm}{\raggedright $U[30,779]$\,d on the PTF and LSST-like cadences} &
  \parbox[t]{3.3cm}{\raggedright $U[0,T/1.5]$ with $T$ the PTF baseline, also for the LSST-like set (their footnote~3)} &
  \parbox[t]{5.6cm}{\raggedright on PTF, keeping $P\le T/1.5$ closes 1.4 of the 4.9 points; at LSST-like cadence, in the period bin with the lowest recovery, our recovery is 2.8 (sinusoid) and 11.5 (sawtooth) times theirs} &
  \parbox[t]{1.5cm}{\raggedright \ref{sec:ptf-lch26}, \ref{sec:periods}, \ref{app:periods}} \\[3pt]
\parbox[t]{2.2cm}{\raggedright DRW amplitude} &
  \parbox[t]{3.3cm}{\raggedright $\sigma$ from Equation~\eqref{eq:sigma}, with $\mathrm{SF}_\infty=\sqrt2\,\sigma$} &
  \parbox[t]{3.3cm}{\raggedright the $\sigma$ of their Figure~3 lies 1.70 times above that of our high-mass systems (ratio of medians)} &
  \parbox[t]{5.6cm}{\raggedright LSST-like sinusoid ratio 3.4; predicted ratio 2.7 with our $\sigma$ scaled up to that of their Figure~3, 3.0 for the $\sqrt2$ alone} &
  \parbox[t]{1.5cm}{\raggedright \ref{sec:amplitude}} \\[3pt]
\parbox[t]{2.2cm}{\raggedright Injected amplitude} &
  \parbox[t]{3.3cm}{\raggedright $A\sim U[0,0.5)$\,mag} &
  \parbox[t]{3.3cm}{\raggedright the same; their $A/\sigma$ axis admits two readings} &
  \parbox[t]{5.6cm}{\raggedright in the $A/\sigma$ bins $[1.0,1.5)$, $[1.5,2.0)$ and $[2.0,2.5)$, our LSST-like sinusoid recovery is 53.2, 84.7 and 98.7 per cent against their 12.8, 23.1 and 37.8 per cent, 2.6 to 4.2 times theirs; above $A=0.45$\,mag, 142 of 142 sinusoids are recovered} &
  \parbox[t]{1.5cm}{\raggedright \ref{sec:snr}} \\[3pt]
\parbox[t]{2.2cm}{\raggedright Rest wavelength $\lambda_{\rm RF}$} &
  \parbox[t]{3.3cm}{\raggedright $6516.05/(1+z)$\,\AA} &
  \parbox[t]{3.3cm}{\raggedright the R band ``in the rest frame of the source''} &
  \parbox[t]{5.6cm}{\raggedright not run; a fixed 6516.05\,\AA\ lowers $\sigma$ about 1.4 times, which would be expected to raise recovery} &
  \parbox[t]{1.5cm}{\raggedright \ref{sec:drw}} \\[3pt]
\parbox[t]{2.2cm}{\raggedright Frame of $\tau$} &
  \parbox[t]{3.3cm}{\raggedright rest-frame $\tau$, not multiplied by $(1+z)$} &
  \parbox[t]{3.3cm}{\raggedright not stated; their Figure~3 matches $\tau(1+z)$} &
  \parbox[t]{5.6cm}{\raggedright multiplying $\tau$ by $(1+z)$ raises recovery on every cadence, most on PTF} &
  \parbox[t]{1.5cm}{\raggedright \ref{app:tauz}} \\[3pt]
\parbox[t]{2.2cm}{\raggedright DRW light-curve generator} &
  \parbox[t]{3.3cm}{\raggedright recursion of Equation~\eqref{eq:ou}} &
  \parbox[t]{3.3cm}{\raggedright \citet{TimmerKoenig1995} with \texttt{astroML}, daily over the observed baseline, then sampled; we read this as unpadded} &
  \parbox[t]{5.6cm}{\raggedright on one system, PTF recovery is \prov{19.2} points higher with an unpadded Timmer \& Koenig generator (series no longer than the light curve) than with Equation~\eqref{eq:ou}, and the same with a padded one (longer series cut to length); LSST-like moves by at most \prov{2} points} &
  \parbox[t]{1.5cm}{\raggedright \ref{app:generator}} \\[3pt]
\parbox[t]{2.2cm}{\raggedright Per-epoch error on PTF} &
  \parbox[t]{3.3cm}{\raggedright 0.05\,mag at every exposure} &
  \parbox[t]{3.3cm}{\raggedright the PTF uncertainty of each epoch, one epoch per night} &
  \parbox[t]{5.6cm}{\raggedright the rise from daily sampling stays at \prov{+10} to \prov{+20} points for 0.01 to 0.15\,mag} &
  \parbox[t]{1.5cm}{\raggedright \ref{app:errors}} \\[3pt]
\parbox[t]{2.2cm}{\raggedright PTF cadences} &
  \parbox[t]{3.3cm}{\raggedright 49 quasars from \citet{Charisi2016}, a mean of 57 nights, repeat exposures kept} &
  \parbox[t]{3.3cm}{\raggedright one epoch per night; about 98 nights implied by their 12-d mean spacing} &
  \parbox[t]{5.6cm}{\raggedright merging the repeat exposures of each night into one epoch \prov{$-1.85$} points; nights rebuilt at the LCH26 spacing inside the observing seasons instead of across the baseline \prov{$-20.6$} points; rise from daily sampling over these merged and rebuilt cadences \prov{+1.6} to \prov{+28.3} points} &
  \parbox[t]{1.5cm}{\raggedright \ref{sec:daily}, \ref{app:nights}} \\[3pt]
\parbox[t]{2.2cm}{\raggedright Mass and magnitude source} &
  \parbox[t]{3.3cm}{\raggedright TNG100-1 masses with DR7 $M_i$} &
  \parbox[t]{3.3cm}{\raggedright 12,400 DR7 quasars} &
  \parbox[t]{5.6cm}{\raggedright reweighting to the DR7 mass distribution within $[8,10)$ moves recovery by at most 0.6 (sinusoid) and 1.4 (sawtooth) points; the rest of the mismatch is untested} &
  \parbox[t]{1.5cm}{\raggedright \ref{sec:lsst-lch26}, \ref{app:ptfdiff}} \\[3pt]
\bottomrule
\end{tabular*}
\notejustify
\tablenotes{Points are percentage points of significance-only sinusoid
recovery in the high-mass range, $\log M_1\in[8,10)$, unless the row says
otherwise; daggered values come from runs with fewer than $10^5$ null light
curves per system. The rise from daily sampling is the recovery with one epoch
per day over a PTF baseline minus the recovery on the observed cadence
(per-epoch error row) or on a version with merged exposures or rebuilt nights
(PTF cadences row). The LCH26 night count follows from their mean spacing of
about 12\,d over their 3.2-yr (1169-d) mean baseline, 97.4 intervals and so
about 98 nights; they give no epoch count.}
\end{table}

\section{Discussion}
\label{sec:discussion}

\subsection{Masses of the binaries a search recovers}
\label{sec:mass-range}

An archival search counts binaries across the whole mass function, and about half of the TNG100-1 mergers lie below $10^8\,M_\odot$, where the sinusoid recovery on the archival cadences with magnitudes drawn from DR7 is lowest; at LSST-like cadence the sinusoid recovery is above 60 per cent in every bin (Table~\ref{tab:perbin}). Because DR7 is flux limited, the magnitudes drawn at low mass are brighter than typical for that mass, so the low-mass recovery fractions are likely too high for a deeper survey, by an amount we have not measured. Host-galaxy light, which we do not model, adds a further caveat at low mass, because \citet{Davis2026} find that host-galaxy contamination ``will increase false-positive rates and decrease binary parameter recovery rates'', most for lower-mass, lower-luminosity binaries.

\subsection{Cadence and mass compared}
\label{sec:cadence-mass}

A forecast that carries these fractions to another survey changes the cadence, and one that carries them to another population changes the mass distribution, so it matters which of the two moves recovery more. The PTF and LSST-like runs share the injected-period range and the recovery definitions, so that pair is the one that isolates what cadence buys. The pair is unpaired in a way theirs is not: the two runs share 1853 of their 3000 systems and draw injection parameters and DRW realisations independently, while LCH26 inject ``the exact same parameters \ldots\ and even the same noise realization'' across their simulations. The LCH26 cadence contrast is a matched-pairs measurement and ours is not, which costs precision on the cadence comparison and does not bias it. In the high-mass range the sinusoid recovery climbs by 37.1 points from the PTF value of Section~\ref{sec:ptf-lch26} to the LSST-like value of Section~\ref{sec:lsst-lch26}. Per-epoch photometric precision travels with cadence across the two, 0.05\,mag on PTF against 0.01\,mag on the LSST-like specification, so the climb measures the whole cadence: more epochs, a longer baseline, and smaller errors on each point. The 37.1-point climb is about the same as the 36-point rise of sinusoid recovery from $\log M_1=6$ to 10 on the Stripe~82 cadence with magnitudes drawn from DR7 (Section~\ref{sec:mass}), with two limits on the comparison: the climb includes the smaller per-epoch error at LSST-like cadence, and 16 of the 36 points, nearly half, come from the step out of the lowest mass bin, whose magnitudes are extrapolated. Within these limits cadence and mass move recovery by comparable amounts.

\subsection{Limitations}
\label{sec:limitations}

Several limits carry into the comparison, with different signs. The effect of $N_{\rm MC}$ is measured twice: raising $N_{\rm MC}$ from $10^4$ to $10^5$ took the high-mass sinusoid recovery on Stripe~82 from \prov{44.5} to 37.0 per cent and lowered the PTF value by a similar amount (Table~\ref{tab:ptf} and Section~\ref{sec:daily}), both drops of \prov{7} to \prov{10} points. The two runs use different system draws, so part of each shift is sampling, but the direction and rough size line up, and a run with fewer null light curves reads high unless nearly every injection is already recovered. The null light curves use each system's true DRW parameters, as in LCH26, while a search of real quasars must estimate them, so the fractions describe the method with known noise properties.

The PTF cadences are 49 \citet{Charisi2016} candidates, selected for detectable periodicity, so their baselines skew long, 3.9\,yr against the 3.2\,yr LCH26 quote, but on sampling they run the other way: they hold fewer nights than the LCH26 light curves (Section~\ref{sec:daily}), and that lowers our PTF recovery. Nineteen of the 49 cadences fall below the 50-night cut \citet{Charisi2016} apply, a shortfall that traces to our archive extraction, since every quasar in their analysed sample clears that cut by construction.

The magnitude draw maps the DR7 virial-mass relation onto TNG100-1 true masses. Virial masses carry a scatter that a true mass does not, and we have not propagated that scatter into the amplitude slope, so the flattening near $-0.06$ (Equation~\eqref{eq:slope}) is a central-value statement. The injected periods, at most 779\,d (1100\,d on Stripe~82) in the observed frame, correspond to gravitational-wave frequencies of at least 21\,nHz whether the optical period is the orbital period or a few times it, at and above the upper end of the band in which pulsar timing arrays measure the background \citep{Agazie2023evidence}, so the fractions bear only indirectly on the binaries that dominate that background. The TNG100-1 merger catalogue is a simulated population with its own selection, and the merger times need not coincide with the epochs at which real binaries shine as luminous quasars. A TNG100-1 black hole merger is also a kiloparsec-scale event, not a resolved subparsec binary, and the delay to true coalescence runs to gigayears and stays uncertain \citep{KelleyBlechaHernquist2017}.

\subsection{Comparison with other recovery studies}
\label{sec:other-studies}

\citet{Bertassi2026} reanalysed the LCH26 light curves with their Gaussian-process model, at Bayes-factor thresholds set from the false-positive rate LCH26 assume ($\log_{10}B=5$ on the PTF-like and idealised cadences and $\log_{10}B=3$ at LSST-like cadence, in the note to their Table~4), and ordered the cadences as our periodogram runs do (Figure~\ref{fig:vs_lch26}). In the X-ray band, daily MAXI and RXTE/ASM monitoring of the brightest active galactic nuclei raised the X-ray completeness of a cross-band periodicity search 5 to 8 times over the sparse Swift-BAT cadence \citep{Akbari2026}, the same direction as our result for dense sampling. LCH26 state no numerical false-positive rate, and Section~\ref{sec:detection} turns their threshold into a bound per light curve. \citet{Chiesa2026} run periodogram recovery on mock-LSST light curves of a semi-analytic binary population with variability from hydrodynamic simulations, and report recovery above 50 per cent for high-eccentricity binaries ($e>0.6$) and false-alarm probabilities above $10^{-1}$ for their circular binaries. A circular binary is the closest analogue of our sinusoid, so read whole, \citet{Chiesa2026} is supporting evidence for strongly modulated eccentric systems and neutral to adverse for the sinusoid.

\citet{Witt2022} bear on the level from the other side, selecting periodicity with the Bayesian information criterion, with the red noise carried inside the test statistic, and LCH26 summarise their detection rates in CRTS-like and LSST-like simulations as about 60 per cent in both; it is their summary we quote and not a figure we read off. \citet{Robnik2024} put the LSP peak at a factor 20 lower true-positive probability than a Bayes factor, quoted at a false-positive probability of $10^{-3}$, so a Bayesian test reading 60 per cent where an LSP reads less runs in the direction they measure. Our rule is stricter than the point that factor is quoted at (Appendix~\ref{app:fap}), so our LSST-like recovery is not the product of a loose threshold. \citet{Witt2022} do not state their false-positive rate, so none of these true-positive rates are compared at one common false-positive rate. No other periodogram recovery fraction for a full quasar sample at LSST-like cadence has been published: \citet{Chiesa2026} report theirs by eccentricity, and \citet{Witt2022} and \citet{Bertassi2026} report model selection, a different quantity. Two published results bear on the level of our LSST-like recovery, the high false-alarm probabilities \citet{Chiesa2026} find for circular binaries and the 80 per cent ceiling of the LCH26 amplitude curve (Section~\ref{sec:snr}), and neither is dismissed here.

\citet{Kelley2019} found detection efficiency rising with total mass for Illustris binaries under a variability-amplitude threshold, strongly for Doppler-boosted variability (from the orbital motion of the emitting gas) and weakly for hydrodynamic variability (from the circumbinary flow). \citet{Davis2024} report that photometric selection above 0.1\,mag amplitude yields samples dominated by false positives, with a false-positive fraction of 60 per cent for the most luminous quasars. The result cuts against the practical use of the mass trend measured here, since the bright, massive hosts our fractions favour are exactly the population they find hardest to keep clean. Recovery and reliability are different quantities, and our false-alarm probabilities (Appendix~\ref{app:fap}) are one input to reliability, but they are conditional on the DRW plus Gaussian null that calibrates the rule, and \citet{Robnik2024} report that a false-positive rate computed on simulated DRW light curves, as ours is, likely underestimated that of the earlier PTF search; the false-positive fraction of a search of real quasars, the quantity behind the 60 per cent of \citet{Davis2024}, is not measured here. Our injected periods start at 30\,d, so they reach only the long end of the 1 to 70\,d orbital periods at which \citet{Xin2021} forecast 20 to 100 million compact binaries in the LSST quasar catalogue.

\section{Conclusions}
\label{sec:conclusions}

We ran Lomb-Scargle injection and recovery on stratified draws of 3000 systems from the 18,374 TNG100-1 mergers, gave every system a DR7 magnitude at its mass (extrapolated below $10^{6.5}\,M_\odot$), and put observed PTF cadences, one observed Stripe~82 cadence and an LSST-like cadence through the pipeline, to see how far the LCH26 recovery fractions carry across black hole mass. Whether detectability rises or falls with mass comes down to how each system's $M_i$ is assigned: hold the magnitude constant and recovery drops toward the high-mass end, and draw it from DR7 instead and the trend turns over on the archival cadences, so the bright, heavy hosts become the easiest to detect. A forecast for an archival search that applies the recovery of massive systems at every mass therefore overstates recovery at lower mass: on PTF the sinusoid recovery in the mass bins below $10^8\,M_\odot$ is 15 to 29 per cent (Table~\ref{tab:perbin}), under the high-mass value of Table~\ref{tab:ptf}. On the PTF cadences the significance-only sinusoid recovery in the high-mass range, with equal mass weight, comes out 4.9 percentage points under the LCH26 value, about 1.4 of them from injected periods sitting above their own light curve's search range ($P>T/1.5$), which our fixed period range allows and the LCH26 construction does not. The PTF sawtooth lands above theirs, and power spilling onto the harmonics explains 2.6 points of the 4.2-point excess. Even the sparse Stripe~82 cadence recovers 37.0 per cent of sinusoids at high mass, so ground-based data already in hand reach luminous massive hosts.

At LSST-like cadence our recovery above $10^8\,M_\odot$ is about three times the published sinusoid fraction and about forty times the published sawtooth fraction, and most of that difference remains unexplained. LCH26 attribute their own PTF advantage to sparse sampling, so we ran their idealised construction, daily sampling of each of our PTF baselines, inside our pipeline, and it goes the other way: adding epochs over the same PTF baselines did not lower recovery beyond its error in any comparison we made, and an independent Gaussian-process analysis of the LCH26 light curves \citep{Bertassi2026} also recovers more on the denser cadences. If circumbinary accretion produces sawtooth-shaped modulation, an LSP search with amplitudes drawn from $U[0,0.5)$\,mag misses most of it on the archival cadences, where the significance-only sawtooth recovery in the high-mass range is 3.3 per cent on Stripe~82 and 13.5 per cent on PTF, and more than half of it even at LSST-like cadence, which keeps the LCH26 warning in force for the data already in hand.

Table~\ref{tab:perbin} gives the significance-only recovery fraction in each 0.5-dex bin of primary mass from $10^6$ to $10^{10}\,M_\odot$, for both signal shapes, on the PTF, Stripe~82 and LSST-like cadences at $N_{\rm MC}=10^5$, for use as mass-dependent detection efficiencies in forecasts; the lower bins are likely too high for a survey deeper than DR7. Table~\ref{tab:inputs} gives the input-by-input comparison with LCH26, and Section~\ref{sec:daily} the test of dense against sparse sampling at fixed baseline.

\section*{Acknowledgments}
This work uses data from the IllustrisTNG project \citep{Nelson2019, Pillepich2018b, Springel2018, Nelson2018, Naiman2018, Marinacci2018} and observation times from the Palomar Transient Factory \citep{Law2009, Rau2009}. This research has made use of the NASA/IPAC Infrared Science Archive, which is funded by the National Aeronautics and Space Administration and operated by the California Institute of Technology. Funding for the SDSS and SDSS-II has been provided by the Alfred P. Sloan Foundation, the Participating Institutions, the National Science Foundation, the U.S. Department of Energy, the National Aeronautics and Space Administration, the Japanese Monbukagakusho, the Max Planck Society, and the Higher Education Funding Council for England.

This manuscript was prepared and edited with the assistance of Anthropic's Claude \citep{claude2026}. All science, analysis, and conclusions are the authors' own, and all quantitative results are reproducible from the pipeline scripts.

\noindent\textit{Facilities:} Sloan, PO:1.2m, IRSA\par

\noindent\textit{Software:} Python 3.12, astropy 7.2.0 \citep{astropy2013, astropy2018, astropy2022}, numpy 2.4.3 \citep{Harris2020}, scipy 1.17.1 \citep{Virtanen2020}, matplotlib 3.10.8 \citep{Hunter2007}, pandas 2.3.3 \citep{McKinney2010}, joblib 1.5.3, h5py 3.16.0, pyvo 1.8.1\par

\section*{Data Availability}

The TNG100-1 merger catalogue and black hole details (\url{https://www.tng-project.org}), the \citet{Shen2011} DR7 quasar catalogue, the PTF epochs (through IRSA) and the Stripe~82 light curves of \citet{MacLeod2010} (\url{http://faculty.washington.edu/ivezic/macleod/qso_dr7/Southern.html}) are public. The injection-recovery outputs, the 49 PTF cadences, the digitised LCH26 curves and the analysis code are available from the authors on reasonable request.

\appendix

\section{False-Alarm Probability and the Redshift Factor of the Damping Time}
\label{app:calibration}

\subsection{False-alarm probability}
\label{app:fap}

We measured where under the bound $25/(N_{\rm MC}+1)$ of Section~\ref{sec:detection} the rule sits by regenerating, seed for seed, the null light curves of 500 combinations of system and cadence from the three DR7-magnitude runs (200 systems drawn at random and run on both the Stripe~82 and the PTF cadences, and a separate random draw of 100 on the LSST-like cadence), and counting leave-one-out exceedances, each null light curve tested against thresholds rebuilt from the other $N_{\rm MC}-1$. The measured rates per light curve are $2.16\times10^{-4}$ (Stripe~82), $2.20\times10^{-4}$ (PTF) and $2.33\times10^{-4}$ (LSST-like), and 1000 fresh null light curves per system, tested against the original thresholds, read $(2.6\pm0.2)\times10^{-4}$ pooled, consistent with the bound within its error. The shortfall below the bound is cross-bin collision, a single null light curve holding the record in more than one bin at once, and it depends on the cadence: the same 200 systems read $2.16\times10^{-4}$ on the Stripe~82 cadence and $2.20\times10^{-4}$ on the PTF cadences. The measured rates are conditional on our own DRW plus Gaussian null with the true DRW parameters known, so they are not false-alarm probabilities for real quasars, and they apply to the $N_{\rm MC}=10^5$ runs only; for the daggered runs at $N_{\rm MC}=10^4$ the bound is ten times higher.

\subsection{Damping time with and without the factor $(1+z)$}
\label{app:tauz}

LCH26 do not state whether they multiply the rest-frame damping time of Equation~\eqref{eq:tau} by $(1+z)$ to bring it into the observed frame, and we do not (Section~\ref{sec:drw}). We ran both versions on all three cadences: 60 systems per cadence, drawn at random with 7 or 8 per mass bin from the 3000 systems of the DR7-magnitude run on that cadence, each carrying 12 injections per signal shape (the injection of that run and 11 new ones from the same distributions), so 720 injections per shape, cadence and version, with the same injections and noise realisations in either version, at $N_{\rm MC}=1000$. Putting the factor in raises significance-only recovery in every combination: \prov{$+2.50\pm2.07$} points on the LSST-like sinusoid and \prov{$+5.83\pm2.52$} on its sawtooth, \prov{$+5.69$} and \prov{$+7.36$} on PTF, and on Stripe~82 under a point, or about two points with a fresh set of random draws, so leaving the factor out lowers every rate reported here. The errors treat the versions as independent and so overstate the uncertainty of this paired design, and at $N_{\rm MC}=1000$ each difference gives a direction and a rough size, no more. PTF is the most sensitive of the three, which makes this choice a candidate part of the PTF shortfall of Section~\ref{sec:ptf-lch26}, and applying the factor would move our LSST-like recovery further above theirs.

\section{Checks on the Comparisons with Published Recovery}
\label{app:checks}

\subsection{DRW parameters in the published figures}
\label{app:figures}

The two figures LCH26 use to describe their DRW parameters do not settle the amplitude convention. The marginal distributions of $\sigma$ and $\tau$ in their Figure~3 line up with Equations~\eqref{eq:tau} and~\eqref{eq:sigma} only if $\sigma$ is read as 10 raised to the right-hand side of Equation~\eqref{eq:sigma}, that is $\mathrm{SF}_\infty=\sqrt2\,\sigma$, and $\tau$ as the observed-frame $\tau_{\rm RF}(1+z)$. The LCH26 worked example is harder to place, and we could not reproduce it. Figure~2 of LCH26 quotes $\sigma=0.27$\,mag and $\tau=1246$\,d for SDSS J085234.09+330100.7, and that quasar's DR7 entry, $M_i=-26.92$ and $\log_{10}(M_{\rm BH}/M_\odot)=9.40$, returns $\sigma\le0.14$\,mag and $\tau$ of at most about 650\,d under every combination of the two readings, with the published pair needing $\log_{10}(M_{\rm BH}/M_\odot)\approx11$. None of it moves a number in this paper, but it does say their figures and their equations run on different amplitude scales, which is why Appendix~\ref{sec:amplitude} carries both.

\subsection{Amplitude of the DRW noise}
\label{sec:amplitude}

The LCH26 equations and their Figure~3 imply different DRW amplitude scales, and recovery falls as the amplitude rises. Our amplitudes follow Equations~\eqref{eq:tau} and~\eqref{eq:sigma} as written, with $\mathrm{SF}_\infty=\sqrt2\,\sigma$, so the ratios of Section~\ref{sec:lsst-lch26} are the like-for-like figures against the method LCH26 describe. The $\sigma$ distribution their Figure~3 plots sits a factor of 1.70 above that of our high-mass systems, as a ratio of medians: 0.155\,mag for $\sqrt2\,\sigma$ from Equation~\eqref{eq:sigma} over the 104,740 DR7 quasars at $\lambda_{\rm RF}=6516.05\,{\rm \AA}/(1+z)$, the quantity their Figure~3 plots, against 0.091\,mag for $\sigma$ of our 1500 LSST-like high-mass systems. Part of the factor is the $\sqrt2$ between $\sigma$ and $\mathrm{SF}_\infty$ and the rest our equal mass weight against their DR7 sample. To push our own measured amplitude response out to that scale, we measured high-mass LSST-like recovery against $x=A/\sigma$ in 40 bins of equal count, 37 or 38 of the 1500 systems each (Figure~\ref{fig:lsst_amp_period}a shows the curve in ten bins), and predicted recovery for noise scaled up by a factor $s$ as the mean of that curve at $x/s$ over the same systems, interpolated in $\log x$ and held flat beyond the measured range; $s=1$ returns the measured values to within 0.1 points. The prediction puts the sinusoid at 60.9 per cent and the sawtooth at 30.9 per cent at $s=1.70$, ratios of 2.7 and 28 against their values, and taking the $\sqrt2$ alone gives 67.3 and 35.6 per cent, ratios of 3.0 and 32. The sinusoid ratio therefore falls by about a fifth at $s=1.70$ and the sawtooth ratio by about a third, which is how much of each ratio depends on the amplitude reading, and a factor of at least 2.7 on the sinusoid does not.

\subsection{Bounds from the injected period distribution}
\label{app:periods}

Recovery at LSST-like cadence falls with injected period. In five equal-count period bins across the high-mass range (300 injections each, edges set separately for each shape), the significance-only sinusoid recovery falls from 90.7 per cent at 31--185\,d to 62.3 per cent at 637--779\,d, and the sawtooth recovery from 83.3 per cent at 31--191\,d to 12.7 per cent at 629--779\,d.

The sub-30\,d part of their draw works against their rate, not for it. The LCH26 search grid stops at $f_{\rm max}=1/30$ cycles per day, so a signal injected below 30\,d falls outside the searched band and can register only through power at other frequencies. With $P\sim U[0,T/1.5]$ on their 3.2-yr mean baseline that tail holds $30/779=3.9$ per cent of their injections, so total failure there lowers their value by about one point if the tail would otherwise be recovered at their mean rate, and by a few points at the rate of our shortest periods.

The rest of their period distribution is bounded two ways, and neither closes the LSST-like difference. We built their mixture from the 49 PTF baselines, kept our measured recovery inside our period range and set it to zero everywhere outside: the sinusoid recovery is then at least 61.9 per cent on our baselines and 68.2 per cent on the same baselines scaled to their stated 3.2-yr mean, and the sawtooth recovery at least 37.9 and 44.3 per cent. And the worst period bin anywhere inside our range is 62.3 per cent on the sinusoid and 12.7 per cent on the sawtooth, 2.8 and 11.5 times their values.

\subsection{DRW light-curve generator}
\label{app:generator}

A further construction difference is the noise generator. LCH26 build the DRW with the \citet{TimmerKoenig1995} method as implemented in \texttt{astroML} \citep{VanderPlas2012}, drawn daily over a baseline matched to that of the observed light curve and then sampled at the cadence, while we draw ours epoch by epoch with the recursion of Equation~\eqref{eq:ou}, which gives an exact stationary DRW at the observed epochs and involves no padding choice. The \texttt{astroML} routine returns a series of exactly the requested length, so we read their description as the unpadded version below, though they do not mention padding. We compared the recursion with two versions of the Timmer \& Koenig method on one system: $M_{\rm BH}=10^{8.5}\,M_\odot$, $M_i=-23$ and rest wavelength 6516.05\,\AA\ with no redshift, so $\tau=214.3$\,d and $\sigma=0.141$\,mag; sinusoids with $P=300$\,d, $A\sim U[0,0.5)$\,mag and random phase; 3000 injections per cadence and seed with 0.02\,mag errors; $N_{\rm MC}=3000$; the PTF cadence of SDSS J150912.07+204004.6 (102 epochs over 3.19\,yr) and the LSST-like cadence. The two Timmer \& Koenig versions draw the DRW for the injected and the null light curves on a daily grid from the one-sided power spectrum $4\sigma^2\tau/[1+(2\pi f\tau)^2]$ with the zero-frequency term removed, scale Gaussian real and imaginary parts so that the expected variance equals the integral of that spectrum over the grid frequencies, and sample the series at the day nearest each epoch. The unpadded version uses a grid from the first to the last epoch only, so its series wraps around periodically over the baseline and has no power below one cycle per baseline, and it lacks the slow trends a stationary DRW shows within a finite record. The padded version draws a series 100 times longer and keeps a stretch of the light curve's length at a random offset, the usual way to include those trends \citep{Uttley2002,Vaughan2003}.

With one system every injection of a run is tested against the same 25 thresholds, each the maximum over the $N_{\rm MC}$ null light curves, so the recovery fraction moves with the random seed by more than its binomial error. On the PTF cadence the significance-only recovery has a standard deviation of \prov{4.8} points over seeds, several times the binomial error, so we give means over 32 seeds on PTF and over 4 on the LSST-like cadence, with standard errors over seeds. On PTF the significance-only recovery is \prov{$21.6\pm0.8$} per cent with Equation~\eqref{eq:ou}, \prov{$40.8\pm1.0$} per cent unpadded and \prov{$22.3\pm0.6$} per cent padded, so leaving out the padding raises it by \prov{$19.2\pm1.3$} points, and with padding the generator agrees with the recursion. On the LSST-like cadence the three give \prov{75.7} to \prov{77.5} per cent, so the generator does not account for the difference in LSST-like recovery. The generator results are single-system diagnostics; their sizes carry no population weight, and their effect on the order of the cadences is untested.

\subsection{Per-epoch photometric error}
\label{app:errors}

Our PTF runs use one 0.05\,mag error at every epoch, where LCH26 use the per-epoch PTF uncertainties of each light curve. The level is theirs, quoted as the sample mean over the same light curves, and what we drop is the spread between light curves; their 0.05\,mag attaches to a point binned by night, and our PTF cadences spend it on each of 127 exposures instead of each of 57 nights. A mis-specified error would not cancel between a PTF cadence and its daily version, which carries about thirteen times the epochs, so we reran both with 0.01, 0.05 and 0.15\,mag at every epoch, and with one error per cadence spread geometrically from 0.03 to 0.20\,mag across the 49 in the order of Table~\ref{tab:ptfcadences}, standing in for their per-light-curve values, on 160 systems (an evenly spaced subset by catalogue row of 40 per bin of the high-mass systems of the LSST-like draw), with the same injections at every error on a cadence and separately drawn injections on its daily version, at $N_{\rm MC}=600$. The rise in recovery from daily sampling stays between \prov{+10} and \prov{+20} points for the sinusoid and between \prov{+23} and \prov{+33} points for the sawtooth, each with an error of about \prov{5} points paired by system. The test ran where the threshold is low and every rate reads high, so it measures how sensitive the rise is to the assumed error, and the size of the rise comes from Section~\ref{sec:daily}.

\begin{table}[htbp]
\centering
\caption{The 49 PTF cadences: number of epochs, number of distinct nights and baseline $T$ of each quasar's R-band record.}
\label{tab:ptfcadences}
\setlength{\tabcolsep}{3pt}
\begin{tabular*}{\textwidth}{@{\extracolsep{\fill}}lrrrlrrr@{}}
\toprule
Quasar & Epochs & Nights & $T$ (yr) & Quasar & Epochs & Nights & $T$ (yr) \\
\midrule
2QZ J095344.7+010354 & 164 & 61 & 4.06 & SDSS J151243.67+195845.1 & 147 & 67 & 5.05 \\
2QZ J235800.2$-$281429 & 42 & 27 & 2.20 & SDSS J151646.10+221724.7 & 168 & 67 & 5.08 \\
FBQS J150911.2+215508 & 114 & 48 & 3.19 & SDSS J152739.97+413234.6 & 144 & 79 & 5.20 \\
PDS 898 & 130 & 61 & 3.04 & SDSS J152903.11+223623.8 & 138 & 55 & 3.17 \\
PKS 2203$-$215 & 59 & 34 & 2.26 & SDSS J153051.79+503440.1 & 89 & 47 & 3.06 \\
SDSS J005158.83$-$002054.1 & 92 & 49 & 3.14 & SDSS J153251.06+335852.2 & 130 & 62 & 3.04 \\
SDSS J005453.30$-$003258.3 & 92 & 40 & 4.26 & SDSS J155308.65+501436.5 & 120 & 56 & 4.63 \\
SDSS J023050.06+005843.1 & 96 & 36 & 4.25 & SDSS J160322.68+200535.2 & 237 & 99 & 5.03 \\
SDSS J024442.77$-$004223.2 & 249 & 132 & 4.25 & SDSS J160454.57+315733.5 & 174 & 75 & 3.21 \\
SDSS J104648.62+513912.6 & 130 & 53 & 5.14 & SDSS J162634.15+325032.6 & 84 & 48 & 3.04 \\
SDSS J132815.49+361715.9 & 126 & 55 & 3.05 & SDSS J170942.58+342316.2 & 115 & 46 & 2.03 \\
SDSS J133254.51+371735.5 & 103 & 51 & 3.05 & SDSS J171122.67+342658.9 & 106 & 43 & 2.03 \\
SDSS J133840.66+315936.4 & 214 & 92 & 4.10 & SDSS J171617.49+341553.3 & 65 & 33 & 2.03 \\
SDSS J134553.57+334336.0 & 285 & 106 & 4.90 & SDSS J171909.93+344001.3 & 69 & 37 & 2.03 \\
SDSS J134556.16+343224.5 & 238 & 99 & 5.07 & SDSS J212939.60+004845.5 & 98 & 57 & 5.26 \\
SDSS J140929.76+535930.2 & 217 & 105 & 4.67 & SDSS J214036.77+005210.1 & 94 & 57 & 5.26 \\
SDSS J141004.41+334945.5 & 127 & 56 & 5.14 & SDSS J214225.29+001643.2 & 100 & 59 & 5.26 \\
SDSS J141244.09+421257.6 & 94 & 50 & 3.04 & SDSS J214357.98+003349.6 & 85 & 50 & 4.97 \\
SDSS J142339.44+471240.8 & 173 & 59 & 4.66 & SDSS J224008.39+263928.4 & 77 & 42 & 2.38 \\
SDSS J143637.44+090155.5 & 192 & 70 & 5.57 & SDSS J231733.66+001128.3 & 72 & 33 & 3.28 \\
SDSS J145713.26+140334.1 & 168 & 44 & 3.17 & SDSS J232135.73+173916.5 & 75 & 50 & 5.22 \\
SDSS J145859.07+153144.7 & 110 & 38 & 3.17 & SDSS J235928.99+170426.9 & 83 & 52 & 5.29 \\
SDSS J150900.70+175114.3 & 114 & 40 & 3.17 & SDSS J235958.72+003345.3 & 72 & 28 & 5.29 \\
SDSS J150912.07+204004.6 & 102 & 48 & 3.19 & UM 269 & 109 & 53 & 3.16 \\
SDSS J151053.24+240943.3 & 125 & 67 & 5.08 &  & & & \\
\bottomrule
\end{tabular*}
\par\leftskip=0pt\rightskip=0pt\parfillskip=0pt plus 1fil\relax
\tablenotes{Quasar names as in \citet{Charisi2016}. Epochs are distinct exposure times; a night is the integer part of the modified Julian date. The 49 records hold 6207 epochs on 2816 nights.}
\end{table}

\subsection{Distribution of nights within the observing seasons}
\label{app:nights}

We measured the night structure in two parts, on the 2000 systems of the PTF run at $N_{\rm MC}=10^4$, with its seed. Across the whole $[6,10)$ range the daily version reads \prov{68.0} per cent for the sinusoid and \prov{39.1} for the sawtooth, against \prov{39.6} and \prov{14.6} per cent on the observed cadences, and the period check takes the high-mass values of the daily version to \prov{72.6} and \prov{45.3} per cent.

Take the repeat exposures first. The PTF run was repeated with the 49 cadences merged to one epoch per night, at the mean exposure time with the same 0.05\,mag error, on the same systems, the same seed and the same $N_{\rm MC}=10^4$, and recovery went slightly down: \prov{$46.4\pm1.5$} per cent on the sinusoid and \prov{$18.2\pm1.2$} on the sawtooth in the high-mass range, against \prov{49.4} and \prov{21.3} on the observed cadences. The sinusoid difference is \prov{$-3.0$} points with a cluster-bootstrap error of \prov{1.9} (the 49 cadences resampled with replacement 4000 times, both versions together), and against it a 2-point gain is excluded at \prov{2.6} standard errors. A paired test on 120 high-mass systems (30 per bin), with 32 sinusoid injections each (the injection the system received in the PTF run at $N_{\rm MC}=10^4$, and new draws) and $N_{\rm MC}=1500$, makes each merged epoch the mean of that night's simulated exposures, so both versions share injections and noise. Against all exposures it resolves the loss at \prov{$-1.85\pm0.47$} points when the merged epochs are weighted with a 0.05\,mag error and \prov{$-3.05\pm0.41$} when they are weighted with $0.05/\sqrt n$ for $n$ exposures, so it is a small real loss and not a null. Repeat exposures therefore explain neither the gap in PTF sinusoid recovery, which merging would widen, nor the rise from daily sampling. Splitting the \prov{$-3.0$}-point loss of the 2000-system run by injected period puts it at short period, \prov{$-14.7$} points from 30 to 60\,d and \prov{$-11.0$} from 60 to 120\,d, where the sawtooth gains, but these period bins are unpaired between the versions and thin at the short end, 38 and 42 injections in the first, so the \prov{$-14.7$} carries about \prov{9} points of binomial error and the split is a direction and not a measurement.

The merged version is not built to match LCH26. The night deficit is untouched by it, since merging every night to a single epoch is what produces the 57.5 nights, and the cadences with at least 50 nights, the LCH26 selection, still average only 68.5 nights, 30 per cent short of the LCH26 count. Applying their selection criterion is not the same as matching their sampling. Measured against the merged version the rise from daily sampling is \prov{+28.3} points, larger than the \prov{+25.3} against the observed cadences, so merging the repeat exposures, the one difference we could remove, costs recovery and explains none of the PTF difference.

Where the nights sit is the part that moves the number, and it needs a comparator built at their density. We rebuilt the merged cadences twice, taking each up to $N={\rm round}(T/12\,{\rm d})+1$ nights, which is the LCH26 12-day mean spacing applied to our own baselines and lifts the mean from 57.5 to 119.5 nights. The two rebuilds differ only in where the new nights go. The season-preserving version puts them inside the observing seasons of the cadence, so the long gaps survive: it shares the nights among the seasons in proportion to season length (at least one per season, at most the season length in days plus one, the remainder to the largest fractional shares, so a one-night season, of length zero, gets exactly one) and places them inside each season by sorted uniform draws at least 1.0\,d apart. The season-preserving version keeps the first and last nights of the record and the night of each one-night season, and no other original night, with 0.05\,mag on every rebuilt epoch. The whole-baseline version treats the whole baseline as one season, so the gaps are broken. Both versions ran the configuration of the PTF run at $N_{\rm MC}=10^4$, on the same systems and the same injections, cadence for cadence. The season-preserving version returns \prov{$52.5\pm2.0$} per cent on the sinusoid and \prov{$20.9\pm1.3$} on the sawtooth, the whole-baseline version \prov{$73.1\pm1.6$} and \prov{$39.0\pm2.6$}. The two carry identical night counts on identical baselines with identical injections, so what separates them is where the nights fall and nothing else: \prov{$+20.6\pm1.9$} points on the sinusoid (\prov{11.2} cluster-robust standard errors, paired system by system) and \prov{$+18.1\pm1.9$} on the sawtooth.

Density inside the seasons matters too, and less. Going from the 57.5 observed nights to 119.5 inside the same seasons is worth \prov{$+6.1\pm1.9$} sinusoid points (\prov{3.1} standard errors), so of the \prov{+26.7} points between the merged and whole-baseline versions, where the nights fall carries about three quarters and density within the seasons the rest. Above that density the sinusoid stops gaining: the whole-baseline version and the daily version agree to \prov{$0.0\pm3.2$} points on the 249 high-mass systems they share (below). The agreement is a sinusoid statement only, and the sawtooth is the check on it, gaining a further \prov{$+13.3\pm3.4$} points on the same systems when the sampling goes daily.

The seasons are ours and not theirs: they come from each cadence's own record, split wherever the observed nights leave a gap longer than 60\,d, and that 60\,d is our choice. The split gives 4.8 seasons per cadence, a median gap of 267\,d between seasons, a duty cycle (summed season length over baseline) of 0.21, and a mean spacing of 2.7\,d between nights inside a season once the season-preserving version is filled, so this is not 12-day sampling laid down inside a season. Placing the nights also runs into a limit: in 40 of the 233 seasons, 17 per cent, the requested nights take the season to its capacity at the 1.0\,d minimum spacing, so the draw inside those is close to daily. Real PTF records do have season structure: the example light curves in Figure~1 of \citet{Charisi2016} show three to four observing stretches with gaps of about 90 to 220\,d between them, read off the figure; the 5-day cadence their text does state is a survey-wide average, and their \S4.7 calls the sampling irregular and not quasi-periodic. Our seasons are narrower and gappier than that: 76 per cent of our gaps between seasons run past 220\,d, and no published specification stands behind them. The \prov{20.6} points fall as the duty cycle rises, from \prov{27.6} points in the sparsest third of the cadences to \prov{11.4} in the densest, so the size is specific to these cadences: a survey whose observing seasons cover a third of the year or more sits denser than nearly every one of them.

Against these versions, daily sampling raises sinusoid recovery by \prov{+22.2} points (\prov{9.3} standard errors) over the season-preserving cadences, which keep the observed seasons, and by \prov{+1.6} points (\prov{0.8} standard errors) over the whole-baseline ones. The full-sample errors treat the daily run and the PTF run at $N_{\rm MC}=10^4$ as independent, but the two stratified draws (375 and 250 systems per bin) share 399 systems, 249 of them high-mass, on which cadence and $M_i$ agree and only $A$, $P$ and $\phi_0$ differ (Appendix~\ref{app:draws}); every merged and rebuilt version shares the same 249. Paired on them, the rises over the season-preserving and whole-baseline cadences are \prov{$+22.5\pm4.6$} points (\prov{4.9} standard errors) and \prov{$0.0\pm3.2$}, with errors narrower than for 249 independent systems but wider than the full-sample errors, which therefore understate the uncertainty of these rises.

\subsection{Injected periods and sawtooth harmonics}
\label{app:ptfdiff}

Six per cent of the high-mass sinusoid injections of the PTF run and 5.9 per cent of the sawtooth ones sit above their own light curve's ceiling, $P>T/1.5$, where the grid cannot reach the injected frequency at all, and they recover at 17.8 and 4.5 per cent against 41.2 and 14.0 for the rest. Restricted to the injections with $P\le T/1.5$, which is the LCH26 construction, the sinusoid reads 41.2 per cent on significance-only and 39.4 after the period check, and the sawtooth 14.0 and 9.5.

The remaining 3.5 points of the sinusoid gap (Section~\ref{sec:ptf-lch26}) have no single identified cause. The PTF cadences are biased long, a mean baseline of 3.9\,yr against their 3.2, which raises our expected recovery and cannot by itself put us under them. Our high-mass range is not their full DR7 sample, and against their mass distribution that mismatch can push either way; the DR7 reweighting of Section~\ref{sec:sampling} tests it only in mass. Only 49 cadences, all \citet{Charisi2016} candidates, stand in for the PTF sample, a selection difference that the widened error does not cover.

Our off-fundamental sawtooth rate, the share of injections detected with the highest peak more than 15 per cent from the injected period, is 4.5 per cent against their 1.9 (Figure~\ref{fig:detection_anatomy} shows such a detection), a surplus of 2.6 points out of the 4.2-point excess. The other 1.6 points sit on the fundamental itself, 9.0 per cent against their 7.4 with the period check, and we have no account of that part.

The sawtooth behaves differently at LSST-like cadence than on the sparse cadences. In the high-mass range 671 sawtooth injections clear the threshold, and 95.1 per cent of them peak at the injected period to within 15 per cent. On the PTF cadences that share is 66.8 per cent, and LCH26 quote 87.8 per cent for their own LSST-like set. Over the full mass range it is 94.7 per cent of 1139 detections, and the sinusoid equivalent is 98.3 per cent. With 484 epochs over 9.66\,yr the fundamental is resolved and the periodogram nearly always puts its tallest peak there, so the harmonic leakage that lifts the PTF sawtooth count above LCH26 belongs to the sparse PTF cadences, and it costs almost nothing at LSST-like cadence.

\section{PTF Cadences and Random Draws}
\label{app:data}

\subsection{PTF archive query}
\label{app:ptfquery}

The PTF cadences start from the 50 quasars in Table~1 of \citet{Charisi2016}, at the sky coordinates that table gives, including for the quasars without SDSS names. For each quasar we queried the PTF archive of the NASA/IPAC Infrared Science Archive (IRSA) through its table access service, first the PTF object table for all R-band objects (filter identifier 2) within 5\,arcsec of the position. We then took from the PTF light-curve table all epochs of those objects that carry the archive's good-quality flag (goodflag = 1), with no other cut. We combined their observation times, counting times that agree to within $10^{-5}$\,d once and keeping repeat exposures within a night. Table~\ref{tab:ptfcadences} lists the 49 cadences we use.

\subsection{Random draws}
\label{app:draws}

All random draws use the default generator of NumPy. The stratified draw of systems uses seed 42. The noise and injection draws of the system in row $i$ of the merger catalogue (counted from 0, with the 18,374 mergers in the order of the public merger file) are seeded with the first 32-bit word that NumPy's \texttt{SeedSequence} generates from $(42,i)$, and that generator draws, in order, the $N_{\rm MC}$ null light curves, the sinusoid injection (amplitude, period and phase, then its DRW and noise) and the sawtooth injection. Drawing the null light curves uses a number of random numbers proportional to the number of epochs, so the same system on two cadences with different numbers of epochs receives different $A$, $P$ and $\phi_0$, and each run draws its injections independently for each system. The paired test of Appendix~\ref{app:tauz} and the 120-system paired test of Appendix~\ref{app:nights} draw their injections apart from this order: the first injection of each system is the one it received in the run the test starts from, and the others come from a separate generator seeded from its row. In Appendix~\ref{app:tauz} both versions reuse these injections and draw their noise from the same random numbers, and in the 120-system test each merged epoch averages the simulated exposures of its night. In the PTF run the cadence of system $i$ is the first word generated from $(43,i)$, modulo 49, an index into Table~\ref{tab:ptfcadences} (alphabetical by quasar name), whose order the errors of Appendix~\ref{app:errors} also follow. The magnitudes were drawn once for the whole catalogue with seed 20260606 and stored for every run. The LSST-like cadence and each rebuilt PTF cadence of Appendix~\ref{app:nights} are single realisations, fixed by their lists of times and shared by every system that uses them.


\begin{thebibliography}{}
\expandafter\ifx\csname natexlab\endcsname\relax\def\natexlab#1{#1}\fi
\providecommand{\url}[1]{\href{#1}{#1}}
\providecommand{\dodoi}[1]{doi:~\href{http://doi.org/#1}{\nolinkurl{#1}}}
\providecommand{\doeprint}[1]{\href{http://ascl.net/#1}{\nolinkurl{http://ascl.net/#1}}}
\providecommand{\doarXiv}[1]{\href{https://arxiv.org/abs/#1}{\nolinkurl{https://arxiv.org/abs/#1}}}

\bibitem[{K.~N. {Abazajian} {et~al.}(2009){Abazajian}, {Adelman-McCarthy},
  {Ag{\"u}eros}, {et~al.}}]{abazajian2009}
{Abazajian}, K.~N., {Adelman-McCarthy}, J.~K., {Ag{\"u}eros}, M.~A., {et~al.}
  2009, ApJS, 182, 543, \dodoi{10.1088/0067-0049/182/2/543}

\bibitem[{G. {Agazie} {et~al.}(2023{\natexlab{a}}){Agazie}, {Anumarlapudi},
  {Archibald}, {et~al.}}]{Agazie2023evidence}
{Agazie}, G., {Anumarlapudi}, A., {Archibald}, A.~M., {et~al.}
  2023{\natexlab{a}}, ApJL, 951, L8,
  \dodoi{10.3847/2041-8213/acdac6}

\bibitem[{G. {Agazie} {et~al.}(2023{\natexlab{b}}){Agazie}, {Anumarlapudi},
  {Archibald}, {et~al.}}]{Agazie2023smbhb}
{Agazie}, G., {Anumarlapudi}, A., {Archibald}, A.~M., {et~al.}
  2023{\natexlab{b}}, ApJL, 952, L37, \dodoi{10.3847/2041-8213/ace18b}

\bibitem[{K. {Akbari}(2026){Akbari}}]{Akbari2026}
{Akbari}, K. 2026, ApJ, 1009, 184,
  \dodoi{10.3847/1538-4357/ae9d7c}

\bibitem[{{Anthropic}(2026){Anthropic}}]{claude2026}
{Anthropic} 2026, Claude (large language model), \url{https://claude.com};
  Claude Opus~4 and Claude~5 model families; last accessed 2026-10-21

\bibitem[{{Astropy Collaboration} {et~al.}(2022){Astropy Collaboration},
  {Price-Whelan}, {Lim}, {et~al.}}]{astropy2022}
{Astropy Collaboration}, {Price-Whelan}, A.~M., {Lim}, P.~L., {et~al.} 2022,
  ApJ, 935, 167, \dodoi{10.3847/1538-4357/ac7c74}

\bibitem[{{Astropy Collaboration} {et~al.}(2018){Astropy Collaboration},
  {Price-Whelan}, {Sip{\H{o}}cz}, {et~al.}}]{astropy2018}
{Astropy Collaboration}, {Price-Whelan}, A.~M., {Sip{\H{o}}cz}, B.~M., {et~al.}
  2018, AJ, 156, 123,
  \dodoi{10.3847/1538-3881/aabc4f}

\bibitem[{{Astropy Collaboration} {et~al.}(2013){Astropy Collaboration},
  {Robitaille}, {Tollerud}, {et~al.}}]{astropy2013}
{Astropy Collaboration}, {Robitaille}, T.~P., {Tollerud}, E.~J., {et~al.} 2013,
  A\&A,
  558, A33, \dodoi{10.1051/0004-6361/201322068}

\bibitem[{M.~C. {Begelman} {et~al.}(1980){Begelman}, {Blandford}, \&
  {Rees}}]{Begelman1980}
{Begelman}, M.~C., {Blandford}, R.~D., \& {Rees}, M.~J. 1980,
  Nature, 287, 307, \dodoi{10.1038/287307a0}

\bibitem[{L. {Bertassi} {et~al.}(2026){Bertassi}, {Charisi}, {Buscicchio},
  {Rigamonti}, {Runnoe}, \& {Dotti}}]{Bertassi2026}
{Bertassi}, L., {Charisi}, M., {Buscicchio}, R., {et~al.} 2026,
  A\&A, 711, A126,
  \dodoi{10.1051/0004-6361/202558583}

\bibitem[{T. {Bogdanovi{\'c}} {et~al.}(2022){Bogdanovi{\'c}}, {Miller}, \&
  {Blecha}}]{Bogdanovic2022}
{Bogdanovi{\'c}}, T., {Miller}, M.~C., \& {Blecha}, L. 2022,
  Living Reviews in Relativity, 25, 3, \dodoi{10.1007/s41114-022-00037-8}

\bibitem[{M. {Charisi} {et~al.}(2016){Charisi}, {Bartos}, {Haiman},
  {et~al.}}]{Charisi2016}
{Charisi}, M., {Bartos}, I., {Haiman}, Z., {et~al.} 2016, MNRAS, 463, 2145, \dodoi{10.1093/mnras/stw1838}

\bibitem[{M. {Charisi} {et~al.}(2022){Charisi}, {Taylor}, {Runnoe},
  {Bogdanovi{\'c}}, \& {Trump}}]{Charisi2022}
{Charisi}, M., {Taylor}, S.~R., {Runnoe}, J., {Bogdanovi{\'c}}, T., \& {Trump},
  J.~R. 2022, MNRAS, 510, 5929,
  \dodoi{10.1093/mnras/stab3713}

\bibitem[{Y.-C. {Chen} {et~al.}(2020){Chen}, {Liu}, {Liao},
  {et~al.}}]{Chen2020}
{Chen}, Y.-C., {Liu}, X., {Liao}, W.-T., {et~al.} 2020,
  MNRAS, 499, 2245,
  \dodoi{10.1093/mnras/staa2957}

\bibitem[{Y.-J. {Chen} {et~al.}(2024){Chen}, {Zhai}, {Liu},
  {et~al.}}]{ChenYJ2024}
{Chen}, Y.-J., {Zhai}, S., {Liu}, J.-R., {et~al.} 2024,
  MNRAS, 527, 12154,
  \dodoi{10.1093/mnras/stad3981}

\bibitem[{A. {Chiesa} {et~al.}(2026){Chiesa}, {Izquierdo-Villalba}, {Sesana},
  {et~al.}}]{Chiesa2026}
{Chiesa}, A., {Izquierdo-Villalba}, D., {Sesana}, A., {et~al.} 2026,
  A\&A, 707, A89,
  \dodoi{10.1051/0004-6361/202557029}

\bibitem[{M.~C. {Davis} {et~al.}(2024){Davis}, {Grace}, {Trump},
  {et~al.}}]{Davis2024}
{Davis}, M.~C., {Grace}, K.~E., {Trump}, J.~R., {et~al.} 2024,
  ApJ, 965, 34,
  \dodoi{10.3847/1538-4357/ad276e}

\bibitem[{M.~C. {Davis} {et~al.}(2026){Davis}, {Trump}, {Charisi},
  {et~al.}}]{Davis2026}
{Davis}, M.~C., {Trump}, J.~R., {Charisi}, M., {et~al.} 2026,
  ApJS, 282, 30, \dodoi{10.3847/1538-4365/ae2c53}

\bibitem[{D.~J. {D'Orazio} {et~al.}(2013){D'Orazio}, {Haiman}, \&
  {MacFadyen}}]{DOrazio2013}
{D'Orazio}, D.~J., {Haiman}, Z., \& {MacFadyen}, A. 2013,
  MNRAS, 436, 2997, \dodoi{10.1093/mnras/stt1787}

\bibitem[{B.~D. {Farris} {et~al.}(2014){Farris}, {Duffell}, {MacFadyen}, \&
  {Haiman}}]{Farris2014}
{Farris}, B.~D., {Duffell}, P., {MacFadyen}, A.~I., \& {Haiman}, Z. 2014,
  ApJ, 783, 134,
  \dodoi{10.1088/0004-637X/783/2/134}

\bibitem[{M.~J. {Graham} {et~al.}(2015){Graham}, {Djorgovski}, {Stern},
  {et~al.}}]{Graham2015}
{Graham}, M.~J., {Djorgovski}, S.~G., {Stern}, D., {et~al.} 2015,
  MNRAS, 453, 1562,
  \dodoi{10.1093/mnras/stv1726}

\bibitem[{C.~R. {Harris} {et~al.}(2020){Harris}, {Millman}, {van der Walt},
  {et~al.}}]{Harris2020}
{Harris}, C.~R., {Millman}, K.~J., {van der Walt}, S.~J., {et~al.} 2020,
  Nature, 585, 357,
  \dodoi{10.1038/s41586-020-2649-2}

\bibitem[{J.~D. {Hunter}(2007){Hunter}}]{Hunter2007}
{Hunter}, J.~D. 2007, Computing in Science \& Engineering, 9, 90

\bibitem[{{\v Z}. {Ivezi{\'c}} {et~al.}(2019){Ivezi{\'c}}, {Kahn}, {Tyson},
  {et~al.}}]{Ivezic2019}
{Ivezi{\'c}}, {\v Z}., {Kahn}, S.~M., {Tyson}, J.~A., {et~al.} 2019,
  ApJ, 873, 111, \dodoi{10.3847/1538-4357/ab042c}

\bibitem[{L.~Z. {Kelley} {et~al.}(2017){Kelley}, {Blecha}, \&
  {Hernquist}}]{KelleyBlechaHernquist2017}
{Kelley}, L.~Z., {Blecha}, L., \& {Hernquist}, L. 2017, MNRAS, 464,
  3131, \dodoi{10.1093/mnras/stw2452}

\bibitem[{L.~Z. {Kelley} {et~al.}(2019){Kelley}, {Haiman}, {Sesana}, \&
  {Hernquist}}]{Kelley2019}
{Kelley}, L.~Z., {Haiman}, Z., {Sesana}, A., \& {Hernquist}, L. 2019,
  MNRAS,
  485, 1579, \dodoi{10.1093/mnras/stz150}

\bibitem[{B.~C. {Kelly} {et~al.}(2009){Kelly}, {Bechtold}, \&
  {Siemiginowska}}]{Kelly2009}
{Kelly}, B.~C., {Bechtold}, J., \& {Siemiginowska}, A. 2009,
  ApJ, 698, 895, \dodoi{10.1088/0004-637X/698/1/895}

\bibitem[{S. {Koz{\l}owski}(2017){Koz{\l}owski}}]{Kozlowski2017}
{Koz{\l}owski}, S. 2017, A\&A,
  597, A128, \dodoi{10.1051/0004-6361/201629890}

\bibitem[{S. {Koz{\l}owski} {et~al.}(2010){Koz{\l}owski}, {Kochanek},
  {Udalski}, {et~al.}}]{Koz2010}
{Koz{\l}owski}, S., {Kochanek}, C.~S., {Udalski}, A., {et~al.} 2010,
  ApJ, 708, 927,
  \dodoi{10.1088/0004-637X/708/2/927}

\bibitem[{N.~M. {Law} {et~al.}(2009){Law}, {Kulkarni}, {Dekany},
  {et~al.}}]{Law2009}
{Law}, N.~M., {Kulkarni}, S.~R., {Dekany}, R.~G., {et~al.} 2009,
  PASP, 121, 1395, \dodoi{10.1086/648598}

\bibitem[{A. {Lin} {et~al.}(2026){Lin}, {Charisi}, \& {Haiman}}]{Lin2026}
{Lin}, A., {Charisi}, M., \& {Haiman}, Z. 2026, ApJ, 997, 316,
  \dodoi{10.3847/1538-4357/ae29a7}

\bibitem[{T. {Liu} {et~al.}(2019){Liu}, {Gezari}, {Ayers}, {et~al.}}]{Liu2019}
{Liu}, T., {Gezari}, S., {Ayers}, M., {et~al.} 2019,
  ApJ, 884, 36,
  \dodoi{10.3847/1538-4357/ab40cb}

\bibitem[{A.~I. {MacFadyen} \& M. {Milosavljevi{\'c}}(2008){MacFadyen} \&
  {Milosavljevi{\'c}}}]{MacFadyen2008}
{MacFadyen}, A.~I., \& {Milosavljevi{\'c}}, M. 2008, ApJ, 672, 83, \dodoi{10.1086/523869}

\bibitem[{C.~L. {MacLeod} {et~al.}(2010){MacLeod}, {Ivezi{\'c}}, {Kochanek},
  {et~al.}}]{MacLeod2010}
{MacLeod}, C.~L., {Ivezi{\'c}}, {\v Z}., {Kochanek}, C.~S., {et~al.} 2010,
  ApJ, 721, 1014, \dodoi{10.1088/0004-637X/721/2/1014}

\bibitem[{F. {Marinacci} {et~al.}(2018){Marinacci}, {Vogelsberger}, {Pakmor},
  {et~al.}}]{Marinacci2018}
{Marinacci}, F., {Vogelsberger}, M., {Pakmor}, R., {et~al.} 2018,
  MNRAS, 480, 5113, \dodoi{10.1093/mnras/sty2206}

\bibitem[{W. {McKinney}(2010){McKinney}}]{McKinney2010}
{McKinney}, W. 2010, in Proceedings of the 9th Python in Science Conference, 56,
  \dodoi{10.25080/Majora-92bf1922-00a}

\bibitem[{J.~P. {Naiman} {et~al.}(2018){Naiman}, {Pillepich}, {Springel},
  {et~al.}}]{Naiman2018}
{Naiman}, J.~P., {Pillepich}, A., {Springel}, V., {et~al.} 2018,
  MNRAS, 477,
  1206, \dodoi{10.1093/mnras/sty618}

\bibitem[{D. {Nelson} {et~al.}(2018){Nelson}, {Pillepich}, {Springel},
  {et~al.}}]{Nelson2018}
{Nelson}, D., {Pillepich}, A., {Springel}, V., {et~al.} 2018,
  MNRAS, 475, 624, \dodoi{10.1093/mnras/stx3040}

\bibitem[{D. {Nelson} {et~al.}(2019){Nelson}, {Springel}, {Pillepich},
  {et~al.}}]{Nelson2019}
{Nelson}, D., {Springel}, V., {Pillepich}, A., {et~al.} 2019,
  Computational Astrophysics and Cosmology, 6, 2,
  \dodoi{10.1186/s40668-019-0028-x}

\bibitem[{A. {Pillepich} {et~al.}(2018{\natexlab{a}}){Pillepich}, {Nelson},
  {Hernquist}, {et~al.}}]{Pillepich2018b}
{Pillepich}, A., {Nelson}, D., {Hernquist}, L., {et~al.} 2018{\natexlab{a}},
  MNRAS, 475, 648,
  \dodoi{10.1093/mnras/stx3112}

\bibitem[{A. {Pillepich} {et~al.}(2018{\natexlab{b}}){Pillepich}, {Springel},
  {Nelson}, {et~al.}}]{Pillepich2018}
{Pillepich}, A., {Springel}, V., {Nelson}, D., {et~al.} 2018{\natexlab{b}},
  MNRAS, 473, 4077, \dodoi{10.1093/mnras/stx2656}

\bibitem[{A. {Rau} {et~al.}(2009){Rau}, {Kulkarni}, {Law}, {et~al.}}]{Rau2009}
{Rau}, A., {Kulkarni}, S.~R., {Law}, N.~M., {et~al.} 2009,
  PASP, 121, 1334, \dodoi{10.1086/605911}

\bibitem[{J. {Robnik} {et~al.}(2024){Robnik}, {Bayer}, {Charisi}, {Haiman},
  {Lin}, \& {Seljak}}]{Robnik2024}
{Robnik}, J., {Bayer}, A.~E., {Charisi}, M., {et~al.} 2024,
  MNRAS, 534, 1609,
  \dodoi{10.1093/mnras/stae2220}

\bibitem[{Y. {Shen} {et~al.}(2011){Shen}, {Richards}, {Strauss},
  {et~al.}}]{Shen2011}
{Shen}, Y., {Richards}, G.~T., {Strauss}, M.~A., {et~al.} 2011,
  ApJS, 194, 45, \dodoi{10.1088/0067-0049/194/2/45}

\bibitem[{V. {Springel} {et~al.}(2018){Springel}, {Pakmor}, {Pillepich},
  {et~al.}}]{Springel2018}
{Springel}, V., {Pakmor}, R., {Pillepich}, A., {et~al.} 2018,
  MNRAS, 475, 676, \dodoi{10.1093/mnras/stx3304}

\bibitem[{K.~L. {Suberlak} {et~al.}(2021){Suberlak}, {Ivezi{\'c}}, \&
  {MacLeod}}]{Suberlak2021}
{Suberlak}, K.~L., {Ivezi{\'c}}, {\v Z}., \& {MacLeod}, C. 2021,
  ApJ, 907, 96, \dodoi{10.3847/1538-4357/abc698}

\bibitem[{Y. {Tang} {et~al.}(2018){Tang}, {MacFadyen}, \& {Haiman}}]{Tang2018}
{Tang}, Y., {MacFadyen}, A., \& {Haiman}, Z. 2018, MNRAS, 476, 2249, \dodoi{10.1093/mnras/sty423}

\bibitem[{J. {Timmer} \& M. {Koenig}(1995){Timmer} \&
  {Koenig}}]{TimmerKoenig1995}
{Timmer}, J., \& {Koenig}, M. 1995, A\&A, 300, 707

\bibitem[{P. {Uttley} {et~al.}(2002){Uttley}, {McHardy}, \&
  {Papadakis}}]{Uttley2002}
{Uttley}, P., {McHardy}, I.~M., \& {Papadakis}, I.~E. 2002,
  MNRAS, 332, 231, \dodoi{10.1046/j.1365-8711.2002.05298.x}

\bibitem[{J.~T. {VanderPlas} {et~al.}(2012){VanderPlas}, {Connolly},
  {Ivezi{\'c}}, \& {Gray}}]{VanderPlas2012}
{VanderPlas}, J.~T., {Connolly}, A.~J., {Ivezi{\'c}}, {\v Z}., \& {Gray}, A.
  2012, in Proceedings of the 2012 Conference on Intelligent Data
  Understanding (CIDU), 47, \dodoi{10.1109/CIDU.2012.6382200}

\bibitem[{S. {Vaughan} {et~al.}(2003){Vaughan}, {Edelson}, {Warwick}, \&
  {Uttley}}]{Vaughan2003}
{Vaughan}, S., {Edelson}, R., {Warwick}, R.~S., \& {Uttley}, P. 2003,
  MNRAS, 345, 1271,
  \dodoi{10.1046/j.1365-2966.2003.07042.x}

\bibitem[{S. {Vaughan} {et~al.}(2016){Vaughan}, {Uttley}, {Markowitz},
  {et~al.}}]{Vaughan2016}
{Vaughan}, S., {Uttley}, P., {Markowitz}, A.~G., {et~al.} 2016,
  MNRAS,
  461, 3145, \dodoi{10.1093/mnras/stw1412}

\bibitem[{P. {Virtanen} {et~al.}(2020){Virtanen}, {Gommers}, {Oliphant},
  {et~al.}}]{Virtanen2020}
{Virtanen}, P., {Gommers}, R., {Oliphant}, T.~E., {et~al.} 2020,
  Nature Methods, 17, 261, \dodoi{10.1038/s41592-019-0686-2}

\bibitem[{C.~A. {Witt} {et~al.}(2022){Witt}, {Charisi}, {Taylor}, \&
  {Burke-Spolaor}}]{Witt2022}
{Witt}, C.~A., {Charisi}, M., {Taylor}, S.~R., \& {Burke-Spolaor}, S. 2022,
  ApJ, 936, 89, \dodoi{10.3847/1538-4357/ac8356}

\bibitem[{C. {Xin} \& Z. {Haiman}(2021){Xin} \& {Haiman}}]{Xin2021}
{Xin}, C., \& {Haiman}, Z. 2021, MNRAS,
  506, 2408, \dodoi{10.1093/mnras/stab1856}

\end{thebibliography}
\end{document}